\documentclass{article}
\usepackage{piner}
\usepackage[figuresright]{rotating}
\begin{document}

\submitted{}
\title{VSOP and Ground-based VLBI Imaging of the TeV Blazar Markarian 421 at Multiple Epochs}

\author{B. G. Piner \& S. C. Unwin}
\affil{Jet Propulsion Laboratory, California Institute of Technology, 4800 Oak
Grove Dr., Pasadena, CA 91109; glenn@herca.jpl.nasa.gov, unwin@huey.jpl.nasa.gov}

\author{A. E. Wehrle}
\affil{Infrared Processing and Analysis Center, Jet Propulsion Laboratory,
California Institute of Technology 100-22, Pasadena, CA, 91125; aew@ipac.caltech.edu}

\author{P. G. Edwards}
\affil{Institute of Space and Astronautical Science, Sagamihara, Kanagawa 229-8510, Japan;
pge@vsop.isas.ac.jp}

\author{A. L. Fey \& K. A. Kingham}
\affil{U.S. Naval Observatory, Earth Orientation Dept., 3450 Massachusetts Ave.,
Washington D.C. 20392; afey@alf.usno.navy.mil, kak@cygx3.usno.navy.mil}

\begin{abstract}
We present thirty VLBI images of the TeV blazar Markarian 421 (1101+384) at 
fifteen epochs spanning the time range from 1994 to 1997, and at six different frequencies
from 2.3 to 43 GHz.  The imaged observations include
a high-resolution 5 GHz VLBI Space Observatory Programme (VSOP)
observation with the HALCA satellite on 1997 November 14; 
full-track VLBA observations from 1994 April, 1996 November, and 1997 May
at frequencies between 5 and 43 GHz; six epochs of VLBA snapshot
observations at frequencies between 2 and 15 GHz from Radio Reference Frame studies; and
five geodetic VLBI observations at 2 and 8 GHz from the archive of the Washington
VLBI Correlator Facility located at the U.S. Naval Observatory.  The dense
time coverage of the images allows us to unambiguously track components in the parsec-scale jet over
the observed time range.  We measure the speeds of three inner jet components 
located between 0.5 and 5 mas from the core (0.3 to 3 pc projected linear
distance) to be 0.19$\pm$0.27,
0.30$\pm$0.07, and $-$0.07$\pm$0.07~$c$ ($H_{0}=65$ km s$^{-1}$ Mpc$^{-1}$).  
If the sole 43 GHz image is excluded, all measured speeds are consistent with no motion.
These speeds differ from
tentative superluminal speeds measured by Zhang \& B\aa\aa th from three epochs of data 
from the early 1980's.
Possible interpretations of these subluminal speeds in terms of the high Doppler factor demanded by
the TeV variability of this source are discussed.
\end{abstract}

\keywords{BL Lacertae objects: individual (Markarian 421) --- galaxies: active ---
galaxies: jets --- techniques: interferometric --- radio continuum: galaxies ---
radiation mechanisms: non-thermal}

\section{Introduction}
\label{intro}
The closest (z=0.03002) BL Lacertae object Markarian 421 (1101+384)
gained fame in 1992 by becoming the first extragalactic
object to be convincingly detected as a source of very high-energy ($\sim 1$ TeV) $\gamma$-rays 
by the Cerenkov imaging telescope at the Whipple Observatory (Punch et al. 1992).
The $\gamma$-ray spectrum of this source has since been found to extend to energies as
high as 5 TeV (Zweerink et al. 1997; Krennrich et al. 1997).
Mrk 421 has displayed rapid variability at TeV energies
on timescales ranging from days (Buckley et al. 1996; Schubnell et al. 1996; Kerrick et al. 1995) to 
$\sim$15 minutes (Gaidos et al. 1996).  These short timescales require the TeV emission to
be highly beamed in order to avoid high pair-production opacity; the shorter timescales
mentioned above require a Doppler factor $\delta\geq 9$ (Gaidos et al. 1996).  They also constrain the
size of the TeV emission region to be remarkably small, of order 1 to 10 light hours in extent.
Multiwavelength observations of TeV/X-ray flares in Mrk 421 (Buckley et al. 1996;
Takahashi et al. 1996; Macomb et al. 1995, 1996)
have indicated that the TeV flares may be generated by an increase in the maximum energy of
the relativistic electron distribution, causing a synchrotron X-ray flare and a subsequent TeV flare
from inverse Compton scattering of the highest energy electrons off of lower energy photons.
The detection of Mrk 421 at TeV energies, along with the subsequent detection of Mrk 501 (Quinn et al.
1996; Bradbury et al. 1997), and as yet unconfirmed reports of TeV emission from
1ES 2344+514 (Catanese et al. 1998), 
PKS 2155-304 (Chadwick et al. 1999), and 3C 66A (0219+428) (Neshpor et al. 1998)
have spurred a renewed interest in the nonthermal emission mechanisms of blazars and the
origin of the differences between X-ray and radio-selected BL Lac objects (e.g. Stecker, De Jager, 
\& Salamon 1996).  

Despite its prominence at TeV energies, Mrk 421 is not as prominent a blazar at other 
wavelengths.  It has been detected at GeV energies by the EGRET detector on the 
{\em Compton Gamma Ray Observatory} (Sreekumar et al. 1996; Lin et al. 1992), but does not have
a high flux compared to other EGRET blazars and shows no evidence for variability (Mukherjee et al. 1997).
The EGRET detection of Mrk 421 thus does not directly require that the GeV emission be beamed.
There is a detection of a spectral break between the EGRET and Whipple energy ranges, with a
photon spectral index of $-1.58\pm0.22$ being measured at GeV energies by EGRET (Sreekumar et al. 1996),
while a value of $-2.56\pm0.12$ was measured by the Whipple group at TeV energies (Zweerink et al. 1997).
Mrk 421 was the first BL Lac object to be established as an X-ray source (Ricketts, Cooke, \& Pounds 1976).
Subsequent observations showed that the X-ray flux varied on timescales of hours to days
(Buckley et al. 1996; Takahashi et al. 1996; Fink et al. 1991).
Optically, Mrk 421 is identified with a luminous elliptical galaxy with a 
nuclear component that emits highly polarized nonthermal radiation
with no detectable emission lines (Ulrich et al. 1975).  The redshift of 0.03 has been determined
from absorption lines (Ulrich et al. 1975; Margon, Jones, \& Wardle 1978).
The optical variability is discussed by Tosti et al. (1998) and by
Liu, Liu, \& Xie (1997), who suggest a periodicity of 23 years in the optical light curve.
The radio flux density has been monitored with the Mets\"{a}hovi radio telescope
at 22 and 37 GHz (Tosti et al. 1998), and by the University of Michigan
Radio Astronomy Observatory (UMRAO) at 4.8, 8.0, and 14.5 GHz.  Inspection of the UMRAO
online database\footnote{http://www.astro.lsa.umich.edu/obs/radiotel/umrao.html} shows that Mrk 421
is variable and has a flat spectrum, but only rarely has its flux density exceeded 1 Jy.

Structural information on this source has historically
been sparser than the spectral and temporal data discussed above.
The large-scale radio structure has been imaged with the VLA by Xu et al. (1995),
Laurent-Muehleisen et al. (1993), and Antonucci \& Ulvestad (1985).
The high dynamic range images of Laurent-Muehleisen et al. (1993) show Mrk 421 to be dominated by an
unresolved core with a large, diffuse halo of angular diameter $\sim3'$ and very low surface brightness
surrounding the core.  VLBI observations of the parsec-scale structure were made by Zhang \& B\aa\aa th (1990)
(hereafter ZB90) at 5 GHz and Zhang \& B\aa\aa th (1991) (hereafter ZB91) at 22 GHz.  They found Mrk 421 to have 
a core-jet structure on parsec scales, with the jet oriented along a position angle of about $-45\arcdeg$.
Upon comparing 3 epochs of data at 5 GHz from the early 1980's, 
they deduced a possible apparent superluminal motion of two
jet components with a speed of about 2.9~$c$ ($H_{0}=65$ km s$^{-1}$ Mpc$^{-1}$).  
However, the flux densities of the jet components they tracked were
relatively faint (as faint as 1 mJy), and it was not clear whether the structural variations could be 
unambiguously characterized as superluminal motion.  The jet
morphology was studied at single epochs by Edwards et al. (1998) and Giovannini et al. (1998).  
This source has also been imaged as part of VLBI   
surveys by Kellermann et al. (1998), Xu et al. (1995), and Polatidis et al. (1995).
Recently, Marscher (1999) has presented images at high frequencies at multiple epochs.

Accurate measurements of changes in the parsec-scale jet structure imaged with VLBI provide constraints
on the jet kinematics and geometry.  When combined with estimates of the Doppler beaming factor (determined,
for example, from the TeV time variability), the apparent motion of jet components can be used to constrain the
Lorentz factor of the jet and the angle of the jet to the line-of-sight (e.g. Unwin et al. 1994).
VLBI observations are thus an important complement to the $\gamma$-ray observations.  
Although VLBI observations do not yet have the resolution
to image the region producing the TeV $\gamma$-rays (maximum of 0.12 pc resolution for the images in this paper,
vs. a size of order 1,000 times smaller for the TeV production region), they do provide the highest resolution
structural information available, and can image the region immediately downstream.
In this paper we analyze a VLBI Space Observatory Programme (VSOP) observation of Mrk 421 from 1997,
three full-track Very Long Baseline Array\footnote{ 
The National Radio Astronomy Observatory is a facility of the National Science Foundation operated
under cooperative agreement by Associated Universities, Inc.} (VLBA) observations,
six VLBA snapshot observations done as part of a study of the
Radio Reference Frame (Johnston et al. 1995), and five geodetic VLBI 
(Rogers et al. 1983; Clark et al. 1985; Rogers et al. 1993) observations of this source.
We analyze a total of 30 images at 15 epochs from 1994 to 1997 to check for motion of the parsec-scale jet
components.  In $\S$~\ref{obs} we discuss the VLBI observations, in $\S$~\ref{results} the results obtained
from imaging and model fitting, and in $\S$~\ref{discussion} the implications of these results.
We use $H_{0}=65$ km s$^{-1}$ Mpc$^{-1}$ throughout the paper.  
Mrk 421 is close enough that the assumed value of $q_{0}$ has a negligible effect.
At the distance of Mrk 421 (136 Mpc, z=0.03), 1 mas corresponds to a linear distance of 0.64 pc,
and a proper motion of 1 mas/yr corresponds to an apparent speed of 2.2~$c$.

\section{Observations}
\label{obs}
VLBI observations of Mrk 421 from several different sources are studied in this paper.
We observed this source with the VLBA in 1994 April, 1996 November, and 1997 May, and
we observed it again with Space VLBI in 1997 November.
Because we wished to accurately measure motions in the parsec-scale
jet, we supplemented these observations with VLBI observations of Mrk 421 made for astrometric
and geodetic purposes between the years 1994 and 1997.  Table~\ref{obslog} lists all of the VLBI
observations imaged for this paper, in chronological order.  
Each type of observation is discussed in more detail below.

\begin{table*}
\caption{Imaged Observations of Mrk 421}
\label{obslog}
\begin{center}
\begin{tabular}{c l l l c} \tableline \tableline
& & \multicolumn{1}{c}{Observation} & &  Frequencies \\ 
Epoch & \multicolumn{1}{c}{Date} & \multicolumn{1}{c}{Type} & 
Antennas\tablenotemark{a} & (GHz) \\ \tableline
1 & 1994 Apr 22 & Astrophysical & VLBA\tablenotemark{b} & 15,43 \\ 
2 & 1995 Oct 17 & Mark III Geodetic & Al,Fo,G2,Gi,Ko,Mk,Nl,Wt & 2,8 \\ 
3 & 1995 Oct 17 & VLBA Astrometric/Geodetic & VLBA$-$Mk,Nl & 15 \\ 
4 & 1996 Jan 30 & Mark III Geodetic & Al,Fo,G2,Gi,Ko,Wt & 2,8 \\ 
5 & 1996 Mar 26 & Mark III Geodetic & G2,Gi,Ko,Ny,Wt & 2,8 \\ 
6 & 1996 Apr 24 & VLBA Astrometric/Geodetic & VLBA & 2,8,15 \\ 
7 & 1996 Sep 3  & Mark III Geodetic & G2,Gi,Ko,Ny,Wt & 2,8 \\ 
8 & 1996 Nov 21 & Astrophysical & VLBA & 5,22 \\ 
9 & 1996 Dec 10 & Mark III Geodetic & Fo,G2,Gi,Ko,Ny,Wt & 2,8 \\ 
10 & 1997 Jan 31 & VLBA Astrometric/Geodetic & VLBA$+$G2,Gi,Ko,Mc,On,We & 2,8 \\ 
11 & 1997 May 3  & Astrophysical & VLBA & 15,22 \\ 
12 & 1997 May 19 & VLBA Astrometric/Geodetic & VLBA$+$G2,Gi,Ko,Mc,On,We & 2,8 \\ 
13 & 1997 Sep 8  & VLBA Astrometric/Geodetic & VLBA$-$Sc$+$Gi,Ko & 2,8 \\ 
14 & 1997 Nov 14 & Astrophysical & Eb,Gb,HALCA,Jb,Mc,Nt,On,Tr,Wb & 5 \\ 
15 & 1997 Dec 18 & VLBA Astrometric/Geodetic & VLBA$+$G2,Gi,Ko,Mc,Ny,On,We & 2,8 \\ \tableline
\end{tabular}
\end{center}
\tablenotetext{a}{Antenna locations and sizes are as follows:
Al = Algonquin, Ontario; 46 m --- Eb = Effelsberg, Germany; 100 m --- 
Fo = Fortaleza, Brazil; 14 m --- G2 = Green Bank, West Virginia; 20 m --- 
Gb = Green Bank, West Virginia; 43 m ---
Gi = Gilcreek, Alaska; 26 m --- 
HALCA; in orbit; 8 m ---
Jb = Jodrell Bank, UK; 26 m --- Ko = Kokee, Hawaii; 20 m --- 
Mc = Medicina, Italy; 32 m ---
Mk = Mauna Kea, Hawaii; 25 m ---
Nl = North Liberty, Iowa; 25 m --- Nt = Noto, Italy; 32 m ---
Ny = Ny Alesund, Norway; 20 m --- On = Onsala, Sweden; 20 m ---
Sc = St. Croix, US Virgin Islands; 25 m --- Tr = Torun, Poland; 32 m ---
Wb = Westerbork, The Netherlands; 25 m --- We = Westford, Massachusetts; 18 m ---
Wt = Wettzell, Germany; 20 m.}
\tablenotetext{b}{VLBA indicates the standard 10 station VLBA.  Antenna names following a minus sign
indicate VLBA stations not used in that observation.  Antenna names following a plus sign indicate
antennas used in addition to the VLBA.}
\end{table*}

Each of the three proposed VLBA observations consisted of two frequencies in alternating scans
of 13 to 22 minutes duration.  Interleaving of scans in this fashion allows full $(u,v)$ plane
coverage to be obtained at each frequency.  Mrk 421 was observed by the VLBA
on 1994 April 22-23, for a total of 13 hours.  Frequencies of 15 GHz
and 43 GHz were observed in alternating scans, each of length 13 minutes.
Calibration scans were done on the sources OJ287, 3C273, and
3C345 at both frequencies.  These data were used for determining the phase
coherence time and the proper parameters to be used in processing the much
weaker Mrk 421 data.  We observed Mrk 421 again with the VLBA on 1996 November 21.
This observation lasted 14 hours, and frequencies of 5 and 22 GHz were observed in alternating
scans of 22 minutes duration.  A 5 GHz image from this dataset has been previously published
by Edwards et al. (1998).  The final VLBA observation occurred on 1997 May 3-4, when Mrk 421
was observed for 14 hours in 22 minute alternating scans at 15 and 22 GHz.

A VSOP observation of Mrk 421 at 5 GHz was made on 1997 November 14 with the HALCA satellite, seven
telescopes of the European VLBI Network (EVN) (Effelsberg, Jodrell Bank, Medicina,
Noto, Onsala, Torun, and an element of the Westerbork Synthesis Radio
Telescope), and the 140' telescope at Green Bank.  
The data were correlated at the VLBA correlator in Socorro.
The Japanese HALCA satellite was launched in 1997 February and
carries an 8 meter antenna through an elliptical orbit 
with an apogee height of 21,400 km, a perigee height of 560 km, and an orbital
period of 6.3 hours.  The data from the satellite are recorded by a network of ground tracking
stations and subsequently correlated with the data from the participating ground telescopes.
The VSOP system and initial science results are discussed by Hirabayashi et al. (1998).
The observation of Mrk 421 on 1997 November 14 consists of 10 hours of full-track data from
most of the participating ground telescopes, and two hours of HALCA data from the tracking
station at Robledo, Spain, covering the portion of HALCA's orbit from near perigee to near apogee.  
The $(u,v)$ plane coverage of the VSOP observation is shown in Figure 1.
This $(u,v)$ plane coverage results in an elliptical beam with high resolution in the east-west
direction.  The correlated flux density of Mrk 421 on the longest space baselines was about 0.1 Jy,
and fringes were detected from HALCA to the reference antenna (Effelsberg)
on these baselines with a signal-to-noise greater than 20, using a fringe solution interval
of 5 minutes.

\begin{figure*}
\plotfiddle{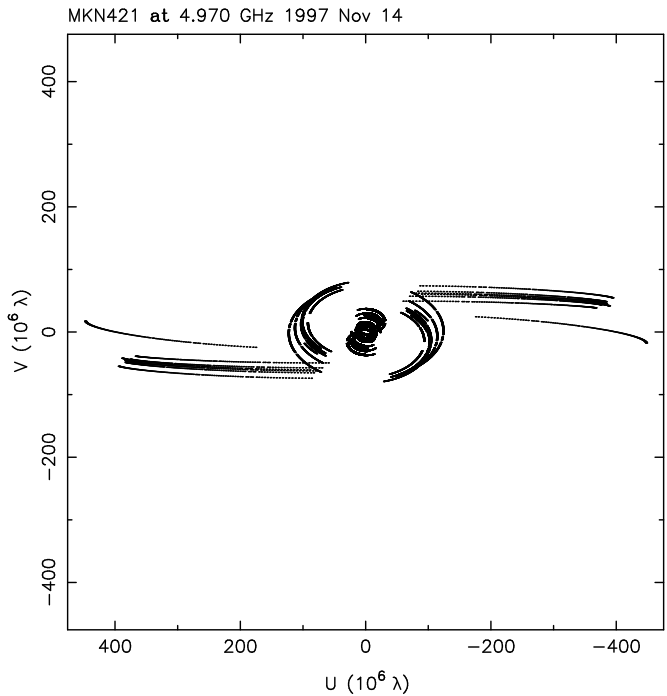}{3.625 in}{0}{50}{50}{-135}{-54}
\caption{$(u,v)$ plane coverage for the VSOP observation of Mrk 421.  The inner arcs are EVN-EVN baselines,
the outer arcs are EVN-Green Bank baselines, and the outer linear segments are the HALCA baselines.}
\end{figure*}

VLBI observations of Mrk 421 have also been done at 
multiple epochs as part of a study of the sources that make up
the Radio Reference Frame (Johnston et al. 1995) (epochs 3, 10, 12, 13, and 15), and once as part  
of a relativity experiment (epoch 6).  The Radio Reference Frame images presented here are the result
of an ongoing program to image the Radio Reference Frame sources on a regular basis.  These images are 
used to monitor sources for variability or structural changes so they can be evaluated for continued
suitability as Radio Reference Frame objects.  The observations are VLBA snapshot observations
at 2, 8, and 15 GHz.  At four epochs 
some geodetic antennas (up to seven) have been used in conjunction with the VLBA.
Fey \& Charlot (1997) and Fey, Clegg, \& Fomalont (1996) discuss the reduction of 
the Radio Reference Frame VLBI observations and present images of some of the Radio Reference Frame sources.
The data for all of these epochs were taken from the Radio Reference Frame Image Database
\footnote{http://www.usno.navy.mil/RRFID} (RRFID).  
Because the RRFID images are produced by running Difmap in an 
automatic mode that can miss low-level extended structure,
all of the images were remade in the standard interactive fashion starting from the original data.

Mrk 421 is one of the sources that is observed regularly in geodetic VLBI experiments.
Five higher quality geodetic VLBI observations of Mrk 421 between the years 1995 and 1997
were selected for analysis in this paper.
The geodetic VLBI data have been taken from the archive of the 
Washington VLBI Correlator Facility located at the U.S. Naval Observatory (USNO).
The USNO geodetic VLBI database and the imaging of geodetic
VLBI data are discussed by Piner \& Kingham (1998, 1997a, 1997b).
The use of geodetic VLBI data allows a large
number of epochs closely spaced in time to be imaged; this is useful in trying to accurately
determine proper motions.  The geodetic observations use
global arrays of antennas, so the images have high resolution.  However, since the observations are
not designed for imaging, the $(u,v)$ plane coverage is often poor and the images are not as high-quality
as those from dedicated imaging arrays.

All of the VLBI observations discussed above were calibrated and fringe-fitted using standard
routines from the AIPS software package, and images were produced using standard CLEAN and
self-calibration procedures from the Difmap software package (Shepherd, Pearson, \& Taylor 1994).

\section{Results}
\label{results}
\subsection{Images}
The VLBI images of Mrk 421 are shown in Figures 2-5.  Figure 2 shows the images
from the VLBA observations on 1994 April 22, 1996 November 21, and 1997 May 3.  
The ten images at 8 and 2 GHz from the
five imaged geodetic VLBI experiments from 1995 October 17 to 1996 December 10 are displayed in Figure 3.  
Figure 4 shows the images from the six epochs of Radio Reference Frame observations of
this source at 15, 8, and 2 GHz from 1995 October 17 to 1997 December 18.  The full-resolution 5 GHz VSOP image
of Mrk 421 from 1997 November 14 is shown in Figure 5$a$.  
In contrast to the other images, which are displayed using
natural weighting (uvweight=0,$-$1 in Difmap), this image is displayed using uniform weighting 
(uvweight=2,0 in Difmap) to preserve the full resolution of the space baselines.  A naturally weighted
VSOP image tends to degrade to the resolution of the ground-only array because of the 
higher weighting of the larger ground antennas.  This is the first Space VLBI image of Mrk 421,
and is the highest resolution 5 GHz image yet obtained of this source.
An image made from the VSOP dataset with the
HALCA baselines removed is shown in Figure 5$b$.  Note the differences in scale between the images 
in Figures 2-5 made from observations at different frequencies and with different arrays.

\begin{sidewaysfigure*}
\plotfiddle{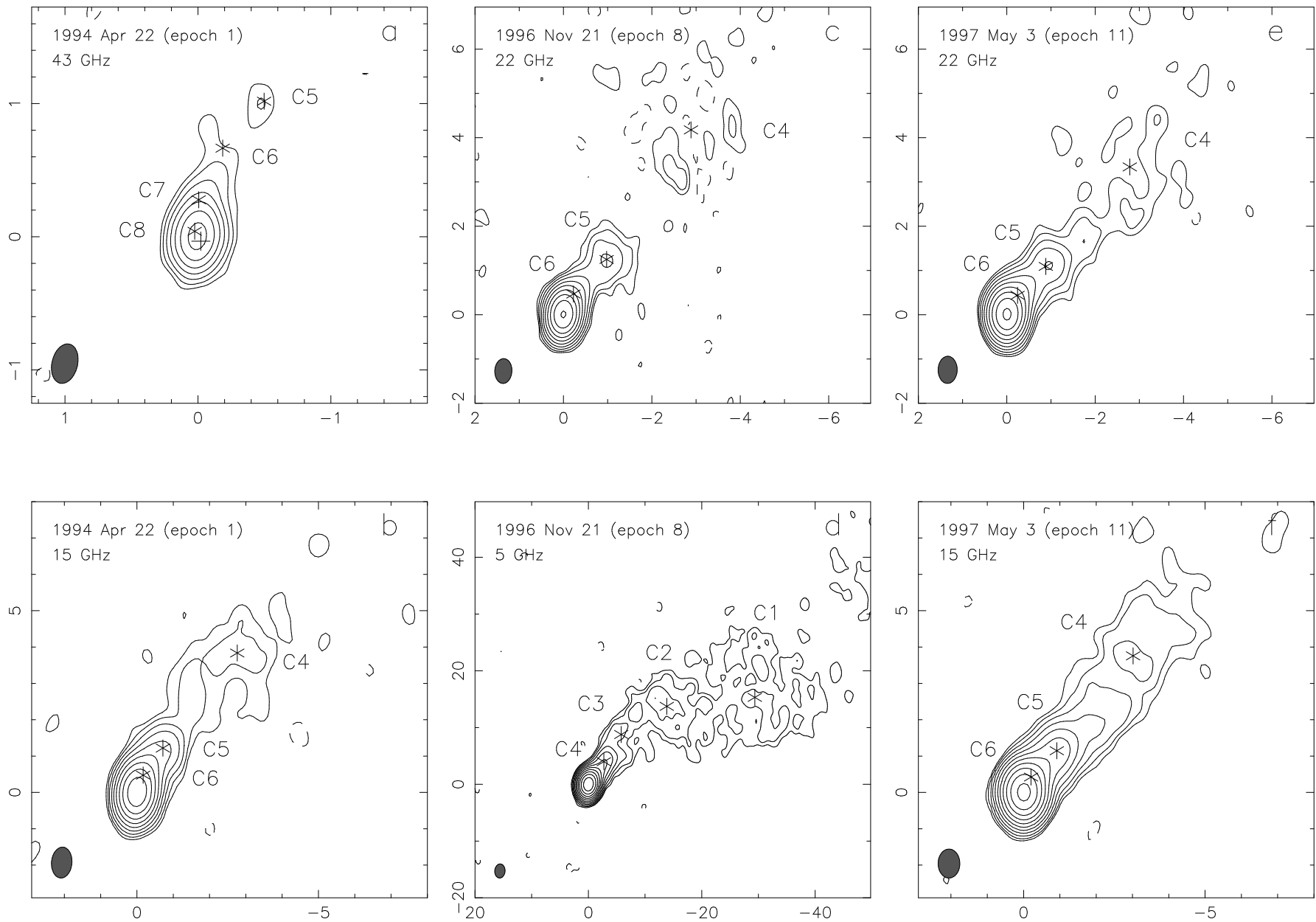}{6.5in}{-90}{90}{90}{-363}{522}
\caption{($a$)-($f$): Full-track VLBA images of Mrk 421 at 43, 22, 15, and 5 GHz
from observations on 1994 April 22, 1996 November 21, and 1997 May 3.
Model-fit component positions
are marked with asterisks.  The model-fit position of the core is marked with a cross
on Figure 2$a$.  Parameters of the images are given in Table 2.}
\end{sidewaysfigure*}

\begin{sidewaysfigure*}
\plotfiddle{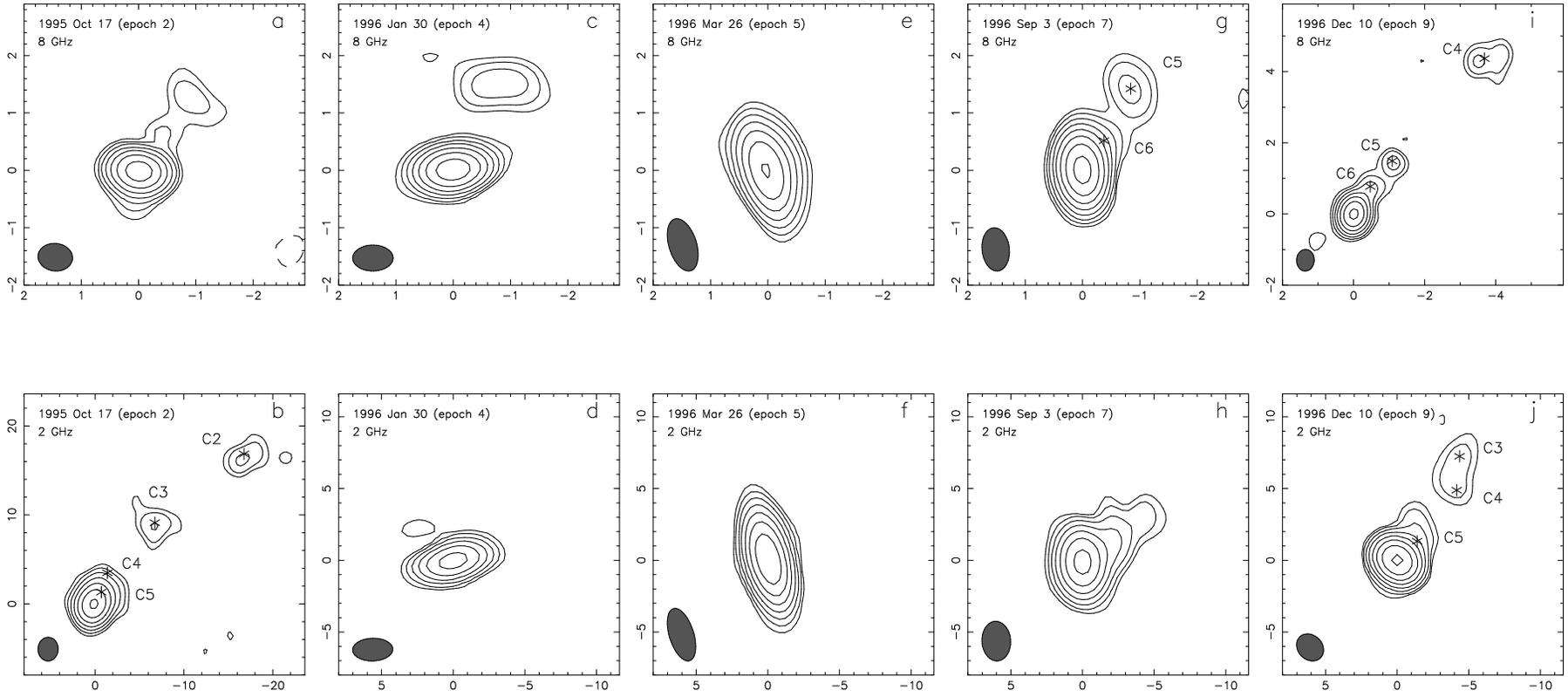}{4.125in}{-90}{90}{90}{-360}{351}
\caption{($a$)-($j$): Geodetic VLBI images of Mrk 421 at 8 and 2 GHz.  All of the 8 GHz images are
5 mas on a side with the exception of $i$ which is plotted on a larger scale.  All of the 2 GHz images
are 20 mas on a side with the exception of $b$ which is plotted on a larger scale.
Model-fit component positions are marked with asterisks
on one image at each scale: Figures 3$b$, $g$, $i$, and $j$.
Parameters of the images are given in Table 2.}
\end{sidewaysfigure*}

\begin{sidewaysfigure*}
\plotfiddle{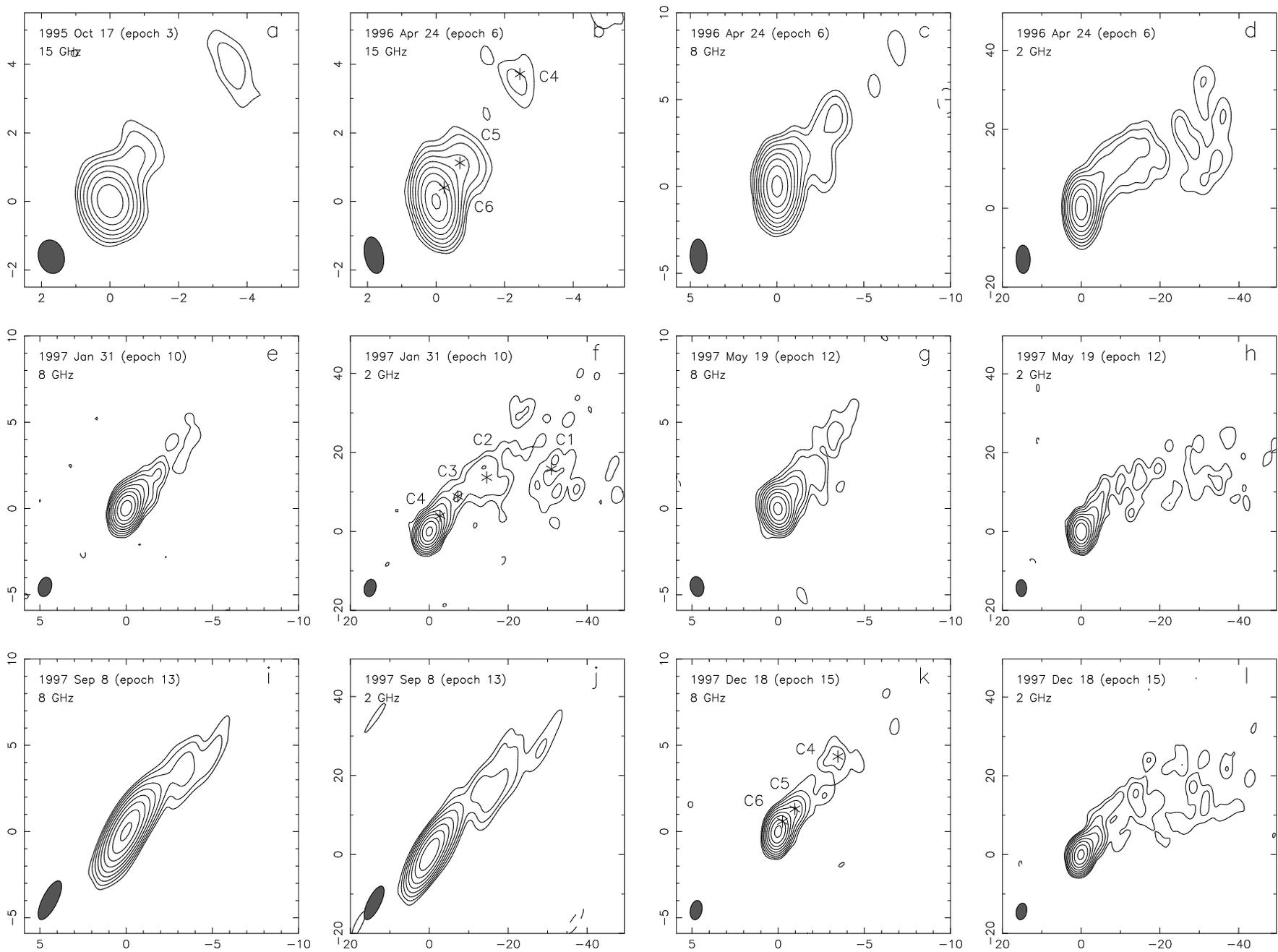}{6.625in}{-90}{90}{90}{-363}{526}
\caption{($a$)-($l$): Images of Mrk 421 from Radio Reference Frame observations at 15, 8,
and 2 GHz.  Model-fit component positions are marked with asterisks
on one image at each frequency: Figures 4$b$, $f$, and $k$.
Parameters of the images are given in Table 2.}
\end{sidewaysfigure*}

\begin{figure*}
\plotfiddle{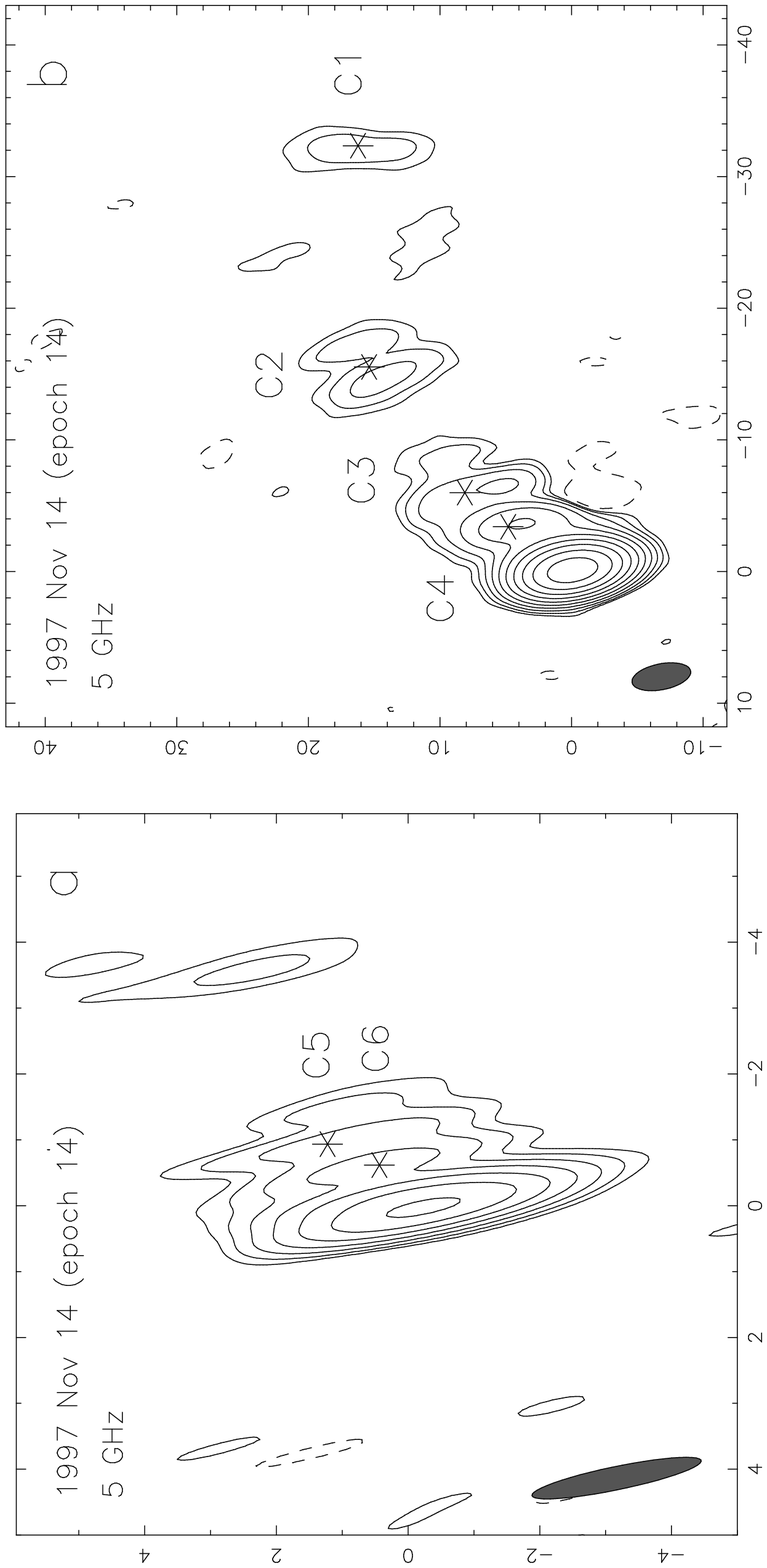}{3.5 in}{-90}{70}{70}{-282}{338}
\caption{($a$)-($b$): Images of Mrk 421 from the 5 GHz VSOP observation on 1997 November 14.  Figure 5$a$
is the full-resolution VSOP image.  Figure 5$b$ shows the image obtained using the ground baselines only.
Model-fit component positions are marked with asterisks.
Parameters of the images are given in Table 2.}
\end{figure*}

In Figures 2-5 all axis labels are
given in milliarcseconds.  The level of the lowest contour in each image has been set equal to
three times the residual rms noise level over the entire image.  Noise contours are not present in some of
the displayed images because the images have been cropped around the source to conserve space.
The relevant parameters for all of these images are listed in Table~\ref{imtab}.  The date, observation
frequency, and epoch number are shown in the upper left hand corner of each image.

\begin{table*}[!t]
\caption{Parameters of the Images}
\label{imtab}
\begin{center}
\begin{tabular}{c c c l c c c} \tableline \tableline
& & & & Peak Flux & Lowest & Contours\tablenotemark{d} \\ 
& & Frequency & & Density & Contour\tablenotemark{c} & (multiples of \\ 
Figure & Epoch\tablenotemark{a} & (GHz) & \multicolumn{1}{c}{Beam\tablenotemark{b}} &
(mJy beam$^{-1}$) & (mJy beam$^{-1}$) & lowest contour) \\ \tableline
2$a$ & 1  & 43 & 0.30,0.19,-13.8 & 192 & 2.22 & 1...2$^{6}$  \\ 
2$b$ &    & 15 & 0.85,0.56,-4.9  & 407 & 0.92 & 1...2$^{8}$  \\ 
2$c$ & 8  & 22 & 0.56,0.39,-2.7  & 404 & 0.75 & 1...2$^{9}$  \\ 
2$d$ &    & 5  & 2.48,1.80,-4.2  & 422 & 0.23 & 1...2$^{10}$ \\ 
2$e$ & 11 & 22 & 0.61,0.43,-1.3  & 268 & 0.47 & 1...2$^{9}$  \\ 
2$f$ &    & 15 & 0.80,0.60,-0.2  & 340 & 0.26 & 1...2$^{10}$ \\ 
3$a$ & 2  & 8  & 0.60,0.48,83.6  & 445 & 4.67 & 1...2$^{6}$  \\ 
3$b$ &    & 2  & 2.67,2.27,-1.4  & 392 & 5.57 & 1...2$^{6}$  \\ 
3$c$ & 4  & 8  & 0.71,0.45,-89.7 & 330 & 3.26 & 1...2$^{6}$  \\ 
3$d$ &    & 2  & 2.80,1.59,-88.5 & 337 & 7.78 & 1...2$^{5}$  \\ 
3$e$ & 5  & 8  & 0.94,0.48,17.3  & 399 & 5.97 & 1...2$^{6}$  \\ 
3$f$ &    & 2  & 3.81,1.71,17.4  & 381 & 3.24 & 1...2$^{6}$  \\ 
3$g$ & 7  & 8  & 0.76,0.48,6.2   & 452 & 2.67 & 1...2$^{7}$  \\ 
3$h$ &    & 2  & 2.77,2.00,3.6   & 411 & 5.01 & 1...2$^{6}$  \\ 
3$i$ & 9  & 8  & 0.61,0.50,-2.5  & 340 & 4.51 & 1...2$^{6}$  \\ 
3$j$ &    & 2  & 2.06,1.70,44.9  & 469 & 3.24 & 1...2$^{7}$  \\ 
4$a$ & 3  & 15 & 0.99,0.75,14.2  & 365 & 3.02 & 1...2$^{6}$  \\ 
4$b$ & 6  & 15 & 1.08,0.54,11.4  & 313 & 1.08 & 1...2$^{8}$  \\ 
4$c$ &    & 8  & 1.98,0.97,2.3   & 298 & 0.90 & 1...2$^{8}$  \\ 
4$d$ &    & 2  & 7.17,3.57,0.9   & 334 & 1.62 & 1...2$^{7}$  \\ 
4$e$ & 10 & 8  & 1.14,0.76,-15.9 & 448 & 1.14 & 1...2$^{8}$  \\ 
4$f$ &    & 2  & 4.44,2.96,-14.1 & 363 & 1.18 & 1...2$^{8}$  \\ 
4$g$ & 12 & 8  & 1.17,0.79,10.7  & 340 & 1.04 & 1...2$^{8}$  \\ 
4$h$ &    & 2  & 4.31,2.66,2.8   & 401 & 1.59 & 1...2$^{7}$  \\ 
4$i$ & 13 & 8  & 2.49,0.86,-26.8 & 405 & 1.37 & 1...2$^{8}$  \\ 
4$j$ &    & 2  & 9.38,3.23,-26.5 & 422 & 1.80 & 1...2$^{7}$  \\ 
4$k$ & 15 & 8  & 1.13,0.68,-11.3 & 363 & 1.12 & 1...2$^{8}$  \\ 
4$l$ &    & 2  & 4.25,2.66,-12.2 & 404 & 1.25 & 1...2$^{8}$  \\ 
5$a$ & 14 & 5  & 2.63,0.35,11.7  & 264 & 3.60 & 1...2$^{6}$  \\ 
5$b$ &    & 5  & 4.48,1.92,11.1  & 357 & 0.44 & 1...2$^{9}$  \\ \tableline
\end{tabular}
\end{center}
\tablenotetext{a}{Epoch identification in Table~\ref{obslog}.}
\tablenotetext{b}{Numbers given for the beam are the FWHMs of the major
and minor axes in mas, and the position angle of the major axis in degrees.}
\tablenotetext{c}{The lowest contour is set to be three times the rms noise
in the full image.}
\tablenotetext{d}{Contour levels are represented by the geometric series 1...2$^{n}$,
e.g. for $n=5$ the contour levels would be $\pm$1,2,4,8,16,32.}
\end{table*}

The higher-resolution images (the 43, 22, and 15 GHz VLBA images, the 8 GHz geodetic images, the 15 and 8 GHz
Radio Reference Frame images, and the 5 GHz VSOP image) all display a similar morphology.  They show Mrk 421
to be core-dominated with a one-sided jet emerging from the core along a position angle of
about $-30$ to $-40\arcdeg$.  This agrees with the morphology observed by other authors at this
angular scale (Marscher 1999; Kellermann et al. 1998; Xu et al. 1995; ZB90).
Note that jet components are not seen in all of the geodetic images.  This is due to the sparse $(u,v)$ plane
coverage and dearth of short baselines in some of the geodetic experiments, and does not necessarily indicate
fading of the components at these epochs.  

The lower-resolution images (the 2 GHz geodetic images, the 2 GHz Radio Reference Frame images, 
the 5 GHz VLBA image, and the 5 GHz 
VSOP ground-baseline only image) show a somewhat different morphology.  They show the jet undergoing a bend
toward a more westerly direction at a distance of about 10 mas from the core.  The jet also becomes broader
and more diffuse at this point (see e.g. Figure 2$d$).
Relatively bright jet emission is detected in these images
out to a distance of about 40 mas, after which there is indication of another bend toward the north.
This is confirmed by the low-resolution images of Giovannini et al. (1998) and Polatidis et al. (1995).
They show the jet becoming extremely broad and decollimated at this point, with diffuse emission extending out 
to distances beyond 100 mas. 

\subsection{Model Fits}
\label{modelfits}
The Difmap model-fitting routine was used to fit elliptical Gaussian components to the visibility data
for each epoch.  The results of this model fitting are given in Table~\ref{mfittab}.  
Occasionally, the axial ratio of an elliptical Gaussian component is not constrained by the data,
and the axial ratio goes to zero during model fitting.  In these cases, one or more circular Gaussians were used
instead.  If the radius of a circular Gaussian became very small then a delta function was used.
There is a problem that arises
when model fitting the VSOP data.  Because the visibilities measured on the space baselines have a lower weight
than those measured on the ground baselines, the model is not constrained to pass through the
space-baseline visibilities.  We addressed this problem by adjusting the weights assigned to the
HALCA visibilities during model fitting.
We increased the weight of HALCA by a factor 
equal to the product of the ratios of the average ground baseline sensitivity to the average space baseline
sensitivity (85 for this observation)
and the number of ground visibilities to the number of space visibilities (12 for this observation).
This applied weight of 1,000 causes the space visibilities to have an effect on the chi-squared
during model fitting equal to that of the ground visibilities.

\begin{table*}
\caption{Gaussian Models}
\label{mfittab}
\begin{center}
\begin{tabular}{c c c r r r r r r r r} \tableline \tableline
& Frequency & Component
& \multicolumn{1}{c}{$S$\tablenotemark{b}} & \multicolumn{1}{c}{$r$\tablenotemark{c}} &
\multicolumn{1}{c}{$\sigma_{r}$\tablenotemark{d}} & \multicolumn{1}{c}{PA\tablenotemark{c}} &
\multicolumn{1}{c}{$\sigma_{{\rm PA}}$\tablenotemark{d}} & \multicolumn{1}{c}{$a$\tablenotemark{e}} &
& \multicolumn{1}{c}{$\Phi$\tablenotemark{f}} \\ 
Epoch\tablenotemark{a} & (GHz) & ID
& \multicolumn{1}{c}{(mJy)} & \multicolumn{1}{c}{(mas)} &
\multicolumn{1}{c}{(mas)} & \multicolumn{1}{c}{(deg)} & \multicolumn{1}{c}{(deg)} 
& \multicolumn{1}{c}{(mas)} & \multicolumn{1}{c}{$b/a$} & \multicolumn{1}{c}{(deg)} \\ \tableline
1  & 43 & Core & 95  & ...   & ...  & ...   & ...  & ...   & ...  & ...   \\ 
   &    & C8   & 116 & 0.07  & ...  & 41.8  & ...  & ...   & ...  & ...   \\ 
   &    & C7   & 14  & 0.29  & ...  & 3.1   & ...  & 0.11  & 1.00 & ...   \\ 
   &    & C6   & 8   & 0.70  & 0.08 & -13.7 & 4.0  & 0.23  & 1.00 & ...   \\ 
   &    & C5   & 6   & 1.13  & 0.07 & -24.8 & 2.5  & ...   & ...  & ...   \\ 
   & 15 & Core & 392 & ...   & ...  & ...   & ...  & 0.18  & 0.53 & -4.9  \\ 
   &    & C6   & 68  & 0.46  & 0.21 & -22.3 & 17.8 & 0.56  & 0.47 & -35.9 \\ 
   &    & C5   & 27  & 1.38  & 0.20 & -31.5 & 6.5  & 0.65  & 0.61 & -71.2 \\ 
   &    & ...  & 4   & 3.05  & ...  & -28.6 & ...  & 1.33  & 0.19 & 10.2  \\ 
   &    & C4   & 23  & 4.69  & 0.39 & -36.1 & 4.0  & 2.80  & 0.63 & -43.8 \\ 
2  & 8  & Core & 444 & ...   & ...  & ...   & ...  & ...   & ...  & ...   \\ 
   &    & C6   & 16  & 0.90  & 0.12 & -11.9 & 9.5  & ...   & ...  & ...   \\ 
   &    & C5   & 14  & 1.61  & 0.12 & -24.1 & 5.3  & ...   & ...  & ...   \\ 
   & 2  & Core & 380 & ...   & ...  & ...   & ...  & 0.56  & 1.00 & ...   \\ 
   &    & C5   & 104 & 1.56  & 0.64 & -36.9 & 21.2 & 1.18  & 1.00 & ...   \\ 
   &    & C4   & 9   & 3.77  & 1.30 & -25.4 & 17.2 & ...   & ...  & ...   \\ 
   &    & C3   & 31  & 11.41 & 1.27 & -37.6 & 6.0  & 1.06  & 1.00 & ...   \\ 
   &    & C2   & 25  & 23.88 & 2.49 & -45.4 & 5.9  & ...   & ...  & ...   \\ 
3  & 15 & Core & 377 & ...   & ...  & ...   & ...  & 0.16  & 0.85 & -40.0 \\ 
   &    & C5   & 23  & 1.27  & 0.22 & -27.4 & 9.7  & 0.42  & 1.00 & ...   \\ 
   &    & C4   & 17  & 5.31  & 0.41 & -43.1 & 5.0  & 0.63  & 1.00 & ...   \\ 
4  & 8  & Core & 333 & ...   & ...  & ...   & ...  & ...   & ...  & ...   \\ 
   &    & C6   & 39  & 0.72  & 0.15 & -47.1 & 11.4 & 0.41  & 1.00 & ...   \\ 
   &    & C5   & 29  & 1.69  & 0.13 & -24.7 & 5.7  & ...   & ...  & ...   \\ 
   & 2  & Core & 349 & ...   & ...  & ...   & ...  & ...   & ...  & ...   \\ 
5  & 8  & Core & 458 & ...   & ...  & ...   & ...  & 0.29  & 0.47 & 24.5  \\ 
   &    & C6   & 26  & 0.48  & 0.12 & -65.3 & 26.0 & ...   & ...  & ...   \\ 
   & 2  & Core & 350 & ...   & ...  & ...   & ...  & ...   & ...  & ...   \\ 
   &    & C5   & 53  & 1.41  & 0.92 & 0.0   & 19.4 & ...   & ...  & ...   \\ 
6  & 15 & Core & 305 & ...   & ...  & ...   & ...  & 0.14  & 0.62 & 42.0  \\ 
   &    & C6   & 43  & 0.44  & 0.22 & -32.6 & 25.5 & 0.32  & 1.00 & ...   \\ 
   &    & C5   & 17  & 1.30  & 0.22 & -32.7 & 9.2  & 0.50  & 1.00 & ...   \\ 
   &    & C4   & 8   & 4.43  & 0.43 & -33.6 & 5.5  & 1.01  & 1.00 & ...   \\ 
   & 8  & Core & 287 & ...   & ...  & ...   & ...  & 0.18  & 1.00 & ...   \\ 
   &    & C6   & 36  & 0.80  & 0.42 & -34.1 & 23.9 & 0.50  & 1.00 & ...   \\ 
   &    & C5   & 15  & 2.13  & 0.41 & -37.6 & 9.8  & 1.18  & 1.00 & ...   \\ 
   &    & C4   & 10  & 5.06  & 0.79 & -41.0 & 8.6  & 1.23  & 1.00 & ...   \\ 
   &    & C2   & 17  & 19.38 & 1.57 & -42.4 & 4.6  & 9.07  & 0.23 & 72.8  \\ 
   & 2  & Core & 339 & ...   & ...  & ...   & ...  & 1.08  & 0.34 & -32.3 \\ 
   &    & C4   & 20  & 4.42  & 3.08 & -35.2 & 30.1 & ...   & ...  & ...   \\ 
   &    & C3   & 16  & 8.96  & 3.07 & -35.5 & 16.0 & 2.33  & 1.00 & ...   \\ 
   &    & C2   & 43  & 18.07 & 5.72 & -43.1 & 17.2 & 11.10 & 0.69 & -73.8 \\ 
   &    & C1   & 38  & 35.06 & 4.63 & -60.7 & 10.6 & 22.95 & 0.45 & -5.6  \\ 
7  & 8  & Core & 463 & ...   & ...  & ...   & ...  & 0.09  & 1.00 & ...   \\ 
   &    & C6   & 14  & 0.61  & 0.16 & -38.4 & 14.6 & ...   & ...  & ...   \\ 
   &    & C5   & 16  & 1.62  & 0.17 & -31.4 & 5.3  & 0.27  & 1.00 & ...   \\ 
   & 2  & Core & 428 & ...   & ...  & ...   & ...  & 0.69  & 0.23 & 3.9   \\ 
   &    & C5   & 47  & 2.60  & 0.55 & -56.5 & 14.0 & ...   & ...  & ...   \\ 
   &    & C4   & 20  & 5.28  & 1.11 & -57.0 & 13.9 & ...   & ...  & ...   \\ 
8  & 22 & Core & 302 & ...   & ...  & ...   & ...  & 0.10  & 1.00 & ...   \\ 
   &    & C7/8 & 139 & 0.15  & ...  & -11.5 & ...  & 0.14  & 1.00 & ...   \\ 
   &    & C6   & 30  & 0.51  & 0.13 & -27.5 & 11.7 & 0.32  & 1.00 & ...   \\ 
   &    & C5   & 18  & 1.58  & 0.13 & -38.7 & 4.1  & 0.63  & 1.00 & ...   \\ 
   &    & C4   & 12  & 5.07  & 0.26 & -34.8 & 2.5  & 3.20  & 1.00 & ...   \\ 
\end{tabular}
\end{center}
\end{table*}

\begin{table*}
\begin{center}
Table 3-Continued \\
\vspace{0.125in}
\begin{tabular}{c c c r r r r r r r r} \tableline \tableline
& Frequency & Component
& \multicolumn{1}{c}{$S$\tablenotemark{b}} & \multicolumn{1}{c}{$r$\tablenotemark{c}} &
\multicolumn{1}{c}{$\sigma_{r}$\tablenotemark{d}} & \multicolumn{1}{c}{PA\tablenotemark{c}} &
\multicolumn{1}{c}{$\sigma_{{\rm PA}}$\tablenotemark{d}} & \multicolumn{1}{c}{$a$\tablenotemark{e}} &
& \multicolumn{1}{c}{$\Phi$\tablenotemark{f}} \\ 
Epoch\tablenotemark{a} & (GHz) & ID
& \multicolumn{1}{c}{(mJy)} & \multicolumn{1}{c}{(mas)} &
\multicolumn{1}{c}{(mas)} & \multicolumn{1}{c}{(deg)} & \multicolumn{1}{c}{(deg)} 
& \multicolumn{1}{c}{(mas)} & \multicolumn{1}{c}{$b/a$} & \multicolumn{1}{c}{(deg)} \\ \tableline
   & 5  & Core & 399 & ...   & ...  & ...   & ...  & 0.19  & 0.24 & -8.4  \\ 
   &    & C6   & 45  & 1.07  & 0.57 & -40.1 & 25.7 & 0.28  & 1.00 & ...   \\ 
   &    & C5   & 10  & 2.24  & 0.57 & -39.7 & 12.9 & 0.39  & 1.00 & ...   \\ 
   &    & C4   & 15  & 4.56  & 1.15 & -37.5 & 12.5 & 2.85  & 0.49 & -29.5 \\ 
   &    & C3   & 12  & 10.25 & 1.16 & -34.8 & 5.6  & 9.36  & 0.43 & -26.0 \\ 
   &    & C2   & 15  & 19.28 & 2.20 & -46.1 & 6.3  & 8.97  & 0.66 & 36.2  \\ 
   &    & C1   & 26  & 32.99 & 2.00 & -63.1 & 4.0  & 18.34 & 0.77 & 83.3  \\ 
   &    & ...  & 38  & 65.60 & ...  & -55.3 & ...  & 40.59 & 0.78 & -74.8 \\
9  & 8  & Core & 371 & ...   & ...  & ...   & ...  & 0.15  & 1.00 & ...   \\ 
   &    & C6   & 48  & 0.86  & 0.15 & -32.8 & 8.7  & 0.40  & 1.00 & ...   \\ 
   &    & C5   & 30  & 1.80  & 0.14 & -36.9 & 4.3  & ...   & ...  & ...   \\ 
   &    & C4   & 47  & 5.67  & 0.28 & -40.4 & 2.7  & 0.81  & 0.51 & -82.7 \\ 
   & 2  & Core & 466 & ...   & ...  & ...   & ...  & ...   & ...  & ...   \\ 
   &    & C5   & 19  & 1.83  & 0.43 & -49.7 & 15.7 & ...   & ...  & ...   \\ 
   &    & C4   & 12  & 6.31  & 0.85 & -41.2 & 9.3  & ...   & ...  & ...   \\ 
   &    & C3   & 15  & 8.34  & 0.86 & -31.5 & 7.0  & ...   & ...  & ...   \\ 
10 & 8  & Core & 433 & ...   & ...  & ...   & ...  & 0.07  & 1.00 & ...   \\ 
   &    & C6   & 45  & 0.60  & 0.28 & -34.6 & 18.5 & 0.48  & 0.17 & 74.8  \\ 
   &    & C5   & 19  & 1.45  & 0.27 & -38.0 & 8.1  & 0.59  & 0.65 & 47.7  \\ 
   &    & ...  & 5   & 2.76  & ...  & -43.0 & ...  & ...   & ...  & ...   \\ 
   &    & C4   & 11  & 5.11  & 0.54 & -40.0 & 4.7  & 3.04  & 0.62 & -26.9 \\ 
   & 2  & Core & 318 & ...   & ...  & ...   & ...  & ...   & ...  & ...   \\ 
   &    & C5   & 56  & 1.42  & 1.06 & -37.9 & 29.7 & ...   & ...  & ...   \\ 
   &    & C4   & 34  & 4.68  & 2.12 & -37.5 & 19.1 & 5.37  & 0.34 & -13.5 \\ 
   &    & C3   & 4   & 11.33 & 2.09 & -40.7 & 8.3  & ...   & ...  & ...   \\ 
   &    & C2   & 30  & 19.90 & 4.05 & -47.6 & 9.9  & 9.20  & 0.87 & -18.8 \\ 
   &    & C1   & 34  & 34.73 & 3.66 & -63.4 & 6.4  & 25.55 & 0.50 & 30.2  \\ 
   &    & ...  & 58  & 63.80 & ...  & -53.7 & ...  & 43.85 & 0.77 & -11.8 \\
11 & 22 & Core & 179 & ...   & ...  & ...   & ...  & 0.10  & ...  & ...   \\ 
   &    & C7/8 & 111 & 0.13  & ...  & 17.1  & ...  & 0.17  & ...  & ...   \\ 
   &    & C6   & 28  & 0.48  & 0.15 & -28.0 & 13.7 & 0.30  & ...  & ...   \\ 
   &    & C5   & 17  & 1.38  & 0.14 & -38.8 & 5.2  & 0.58  & ...  & ...   \\ 
   &    & ...  & 3   & 2.38  & ...  & -44.9 & ...  & 0.46  & ...  & ...   \\ 
   &    & C4   & 13  & 4.33  & 0.28 & -39.8 & 3.4  & 3.32  & 0.57 & -27.0 \\ 
   & 15 & Core & 336 & ...   & ...  & ...   & ...  & 0.19  & 0.63 & 24.4  \\ 
   &    & C6   & 49  & 0.43  & 0.19 & -28.4 & 20.8 & 0.55  & 0.38 & -54.4 \\ 
   &    & C5   & 22  & 1.43  & 0.18 & -39.8 & 6.8  & 0.66  & 0.62 & -54.1 \\ 
   &    & ...  & 6   & 2.78  & ...  & -45.7 & ...  & 1.14  & 0.57 & -65.4 \\ 
   &    & C4   & 13  & 4.78  & 0.36 & -39.1 & 4.1  & 3.31  & 0.44 & -38.2 \\
12 & 8  & Core & 335 & ...   & ...  & ...   & ...  & 0.17  & 0.67 & -5.5  \\ 
   &    & C6   & 34  & 0.60  & 0.25 & -35.4 & 22.5 & 0.21  & 1.00 & ...   \\ 
   &    & C5   & 23  & 1.49  & 0.24 & -40.4 & 9.9  & 0.69  & 0.62 & -2.6  \\ 
   &    & ...  & 12  & 2.95  & ...  & -45.6 & ...  & 1.46  & 1.00 & ...   \\ 
   &    & C4   & 11  & 5.65  & 0.49 & -37.6 & 5.2  & 2.50  & 0.37 & -47.3 \\ 
   & 2  & Core & 400 & ...   & ...  & ...   & ...  & 0.52  & 0.22 & -37.2 \\ 
   &    & C5   & 33  & 2.54  & 0.84 & -49.9 & 20.4 & 0.94  & 1.00 & ...   \\ 
   &    & C4   & 20  & 5.64  & 1.90 & -34.2 & 16.6 & 1.40  & 1.00 & ...   \\ 
   &    & C3   & 6   & 10.10 & 1.95 & -30.1 & 9.1  & ...   & ...  & ...   \\ 
   &    & C2   & 23  & 17.44 & 3.60 & -41.6 & 11.6 & 13.77 & 0.14 & -70.1 \\ 
   &    & C1   & 78  & 34.48 & 2.94 & -65.5 & 6.8  & 35.03 & 0.38 & -64.6 \\ 
13 & 8  & Core & 399 & ...   & ...  & ...   & ...  & 0.23  & 0.69 & -20.1 \\ 
   &    & C5/6 & 34  & 1.27  & 0.60 & -43.1 & 12.0 & 0.33  & 1.00 & ...   \\ 
   &    & C4   & 22  & 4.95  & 1.19 & -44.3 & 6.4  & 4.00  & 0.25 & -55.4 \\ 
   &    & C2   & 33  & 19.65 & 2.34 & -48.2 & 3.5  & 7.15  & 1.00 & ...   \\ 
\end{tabular}
\end{center}
\end{table*}

\begin{table*}[!t]
\begin{center}
Table 3-Continued \\
\vspace{0.125in}
\begin{tabular}{c c c r r r r r r r r} \tableline \tableline
& Frequency & Component
& \multicolumn{1}{c}{$S$\tablenotemark{b}} & \multicolumn{1}{c}{$r$\tablenotemark{c}} &
\multicolumn{1}{c}{$\sigma_{r}$\tablenotemark{d}} & \multicolumn{1}{c}{PA\tablenotemark{c}} &
\multicolumn{1}{c}{$\sigma_{{\rm PA}}$\tablenotemark{d}} & \multicolumn{1}{c}{$a$\tablenotemark{e}} &
& \multicolumn{1}{c}{$\Phi$\tablenotemark{f}} \\ 
Epoch\tablenotemark{a} & (GHz) & ID
& \multicolumn{1}{c}{(mJy)} & \multicolumn{1}{c}{(mas)} &
\multicolumn{1}{c}{(mas)} & \multicolumn{1}{c}{(deg)} & \multicolumn{1}{c}{(deg)} 
& \multicolumn{1}{c}{(mas)} & \multicolumn{1}{c}{$b/a$} & \multicolumn{1}{c}{(deg)} \\ \tableline
   & 2  & Core & 411 & ...   & ...  & ...   & ...  & 0.37  & 1.00 & ...   \\ 
   &    & C4   & 38  & 4.85  & 4.48 & -44.9 & 23.7 & 1.77  & 1.00 & ...   \\ 
   &    & C3   & 14  & 8.59  & 4.67 & -32.1 & 11.0 & 1.54  & 1.00 & ...   \\ 
   &    & C2   & 44  & 24.46 & 9.05 & -42.9 & 9.5  & 15.29 & 0.40 & -39.2 \\ 
14 & 5  & Core & 367 & ...   & ...  & ...   & ...  & 0.39  & 0.32 & -50.4 \\ 
   &    & C6   & 65  & 0.83  & 0.32 & -50.5 & 34.9 & 0.36  & 1.00 & ...   \\ 
   &    & C5   & 24  & 1.63  & 0.45 & -36.1 & 16.7 & 0.12  & 1.00 & ...   \\ 
   &    & C4   & 21  & 5.51  & 1.62 & -38.7 & 18.3 & 3.29  & 0.39 & 21.0  \\ 
   &    & C3   & 10  & 9.70  & 1.63 & -38.5 & 10.6 & 7.15  & 0.15 & 20.4  \\ 
   &    & C2   & 8   & 21.54 & 2.90 & -46.3 & 10.3 & 6.99  & 0.61 & 56.1  \\ 
   &    & C1   & 4   & 36.00 & 2.18 & -64.1 & 6.9  & 8.28  & 0.46 & 32.0  \\ 
15 & 8  & Core & 347 & ...   & ...  & ...   & ...  & 0.18  & 0.37 & 31.9  \\ 
   &    & C6   & 52  & 0.57  & 0.28 & -27.3 & 17.6 & 0.56  & 0.60 & -51.4 \\ 
   &    & C5   & 23  & 1.56  & 0.26 & -38.8 & 7.2  & 0.49  & 1.00 & ...   \\ 
   &    & ...  & 7   & 3.15  & ...  & -45.7 & ...  & 1.91  & 0.49 & 76.7  \\ 
   &    & C4   & 13  & 5.50  & 0.52 & -39.2 & 4.1  & 1.79  & 0.69 & 69.9  \\ 
   & 2  & Core & 414 & ...   & ...  & ...   & ...  & 0.79  & 0.63 & -41.7 \\ 
   &    & C5   & 30  & 2.83  & 0.97 & -43.8 & 15.6 & 1.09  & 1.00 & ...   \\ 
   &    & C4   & 27  & 5.79  & 1.97 & -41.3 & 15.1 & 2.20  & 0.60 & 64.0  \\ 
   &    & C3   & 6   & 10.46 & 2.08 & -26.9 & 7.6  & 1.56  & 1.00 & ...   \\ 
   &    & C2   & 21  & 17.55 & 3.90 & -43.0 & 10.2 & 13.56 & 0.49 & -48.2 \\ 
   &    & C1   & 92  & 32.87 & 3.36 & -63.9 & 6.5  & 35.25 & 0.64 & -87.8 \\ \tableline
\end{tabular}
\end{center}
\tablenotetext{a}{Epoch identification in Table~\ref{obslog}.}
\tablenotetext{b}{Flux density in millijanskys.}
\tablenotetext{c}{$r$ and PA are the polar coordinates of the
center of the component relative to the presumed core.
Position angle is measured from north through east.}
\tablenotetext{d}{$\sigma_{r}$ and $\sigma_{{\rm PA}}$ are the
estimated errors in the component positions.}
\tablenotetext{e}{$a$ and $b$ are the FWHM of the major and minor axes of the Gaussian
component.}
\tablenotetext{f}{Position angle of the major axis measured from north through east.}
\end{table*}

The time sampling of our epochs is dense enough that we are able to 
identify the same components from epoch to epoch.
We have identified a total of six jet components that are present in images at many epochs.
We refer to these components as C1 to C6 from the outermost component inward.
Components C1 to C3 are beyond about 10 mas from the core and are only visible in the lower-resolution
images.  Component C6 is about 0.5 mas from the core and is seen only in the
high-resolution images, while C4 and C5 are detected in images of varying resolution.
Components are detected interior to C6 in the 43 and 22 GHz images, and are identified as C7 and C8
in Table~\ref{mfittab}; however, this region is more confused and it is not clear how these components
should be deconvolved from the core, or from each other in the 22 GHz images.  
In addition, a faint component is sometimes detected between C4 and
C5, and some of the low-resolution images show an extended component to the northwest of C1.

It was noted by Piner \& Kingham (1998) that formal methods for determining the statistical
uncertainties in model-fit parameters generally underestimate the errors for the geodetic VLBI 
data because these methods do not take account of the systematic differences between arrays
and $(u,v)$ plane coverages from image to image.  These differences can be large for the geodetic
VLBI experiments, and are also expected to be large in this work since we are combining data from
four different kinds of VLBI observations (full-track VLBA, snapshot VLBA, geodetic, and VSOP).
Following Piner \& Kingham (1998), we assign the positional error bars for the brighter, 
more compact components C5 and C6 to be one-quarter of the beam FWHM
(maximum projection along the direction from the core to the component), those for the more diffuse
C3 and C4 to be one-half of the beam FWHM, and we estimate that we can measure the centroids of the extended
emission regions referred to as C1 and C2 only to approximately one FWHM of the synthesized beam.

\subsection{Flux Densities}
The sums of the flux densities of the VLBI model components 
at each epoch at 5, 8, and 15 GHz can be compared with the UMRAO
single-dish light curves at 4.8, 8.0, and 14.5 GHz.  For this analysis, we disregard the 8 GHz geodetic
data, because the absolute flux density calibration of the geodetic experiments is not very precise.
At all three frequencies we find that the flux density of the VLBI models is about 70\% of the
single-dish flux density, indicating that there is substantial extended emission.  This figure
agrees approximately with the ratio of core to extended flux density found by Laurent-Muehleisen et al. (1993) 
for this source.  Only the 8 GHz images have a time sampling dense enough to allow the structure of the
VLBI light curve to be compared with that of the single-dish light curve.  Figure 6 shows that the
sums of the model component flux densities from the five
8 GHz VLBA images roughly track the structure in the single-dish light curve, with both showing a peak in
the emission in early 1997 due to an increase in the flux density of the VLBI core.
The core spectrum is relatively flat between 2 and 22 GHz.  We measure the spectral index of the core region
to be $-$0.14 ($S\propto\nu^{+\alpha}$)
between 2 and 22 GHz for the nearly simultaneous VLBA observations at 2, 8, 15, and 22 GHz
from 1997 May.  This agrees well with the average spectral index of the UMRAO light curves between 4.8
and 14.5 GHz of $-$0.17.  The spectrum of the core steepens beyond 22 GHz; we measure a spectral index
of $-$0.54 for the core region between 15 and 43 GHz from the VLBA observation in 1994 April. 
Because the jet components in Mrk 421 are relatively faint ($\sim$30 mJy),
and because these VLBI observations vary in their sensitivity to extended structure,
comparison of the flux densities of the jet components between different frequencies and epochs is not meaningful.

\begin{figure*}
\plotfiddle{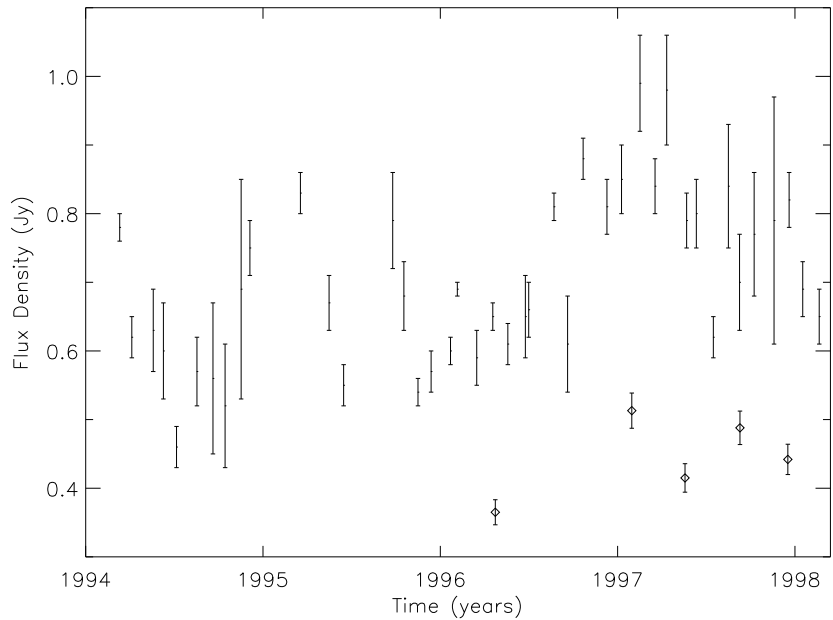}{3.25 in}{90}{50}{50}{200}{-27}
\caption{8.0 GHz light curve of Mrk 421 from the University of Michigan Radio Astronomy Observatory.
The sums of the model component flux densities from the five 8 GHz VLBA images are shown as diamonds.  A 5\%
calibration error is assumed for the VLBI flux densities, based on the corrections given by amplitude
self-calibration.
The sum of the model component flux densities is independent of the model fitting errors on individual
components.}
\end{figure*}

\subsection{Component Positions}
The separations from the core of components C4-C6 over the observed time range are shown in
Figures 7$a$-$c$ respectively.  The error bars shown on these figures have been calculated
as described in $\S$~\ref{modelfits}.
In cases where a component was observed at different frequencies at the same epoch,
a small shift was applied to the times of the plotted positions 
(but not the fitted positions) at different frequencies to avoid 
overlapping error bars.  Only components C4-C6 are shown because they are detected in the
higher-frequency images and have more accurate position measurements.  Components C1-C3 are located
in the outer part of the jet, which is not as easily described by a series of compact components.

\begin{sidewaysfigure*}
\plotfiddle{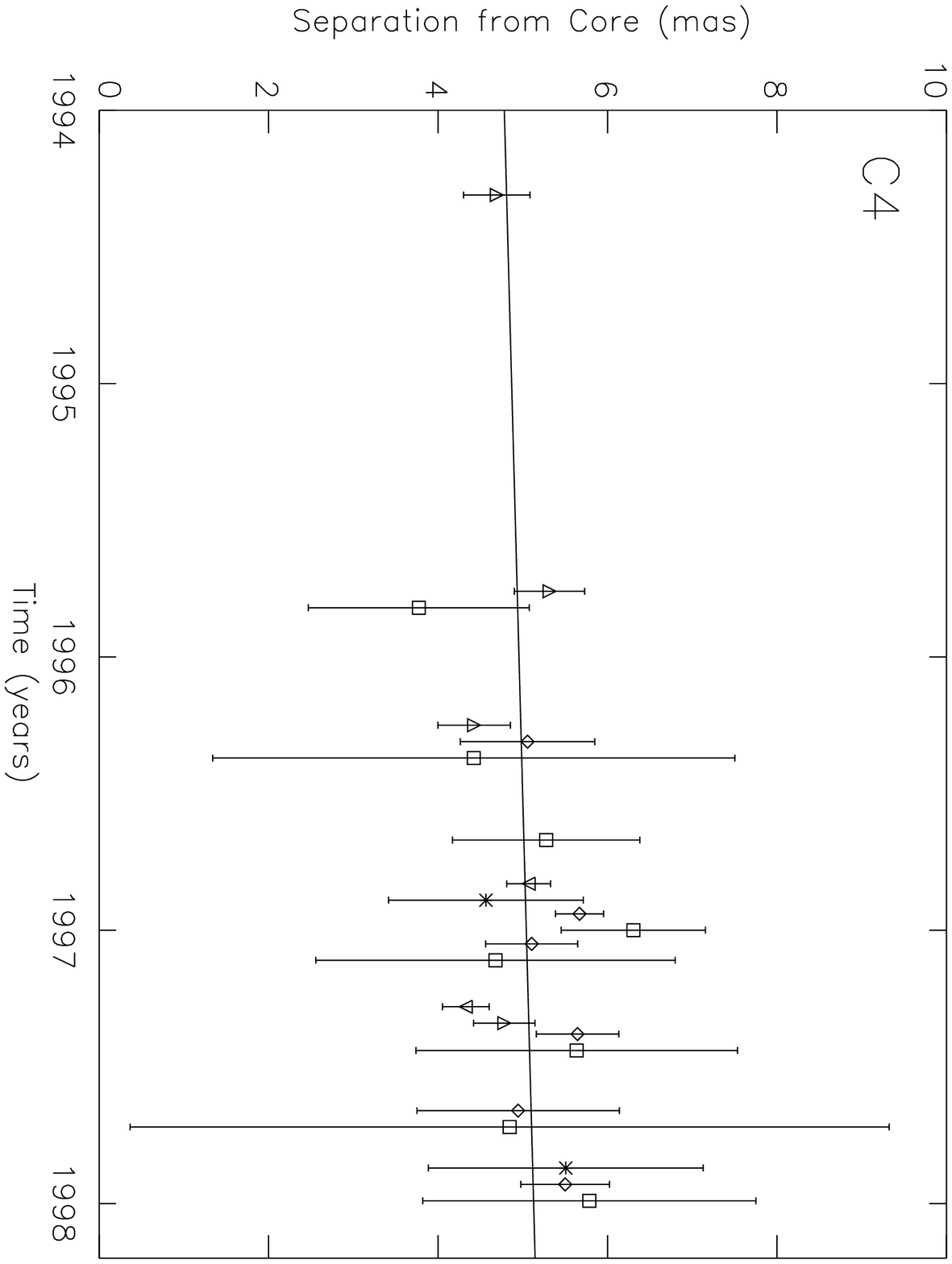}{6.375in}{90}{50}{50}{29}{193}
\plotfiddle{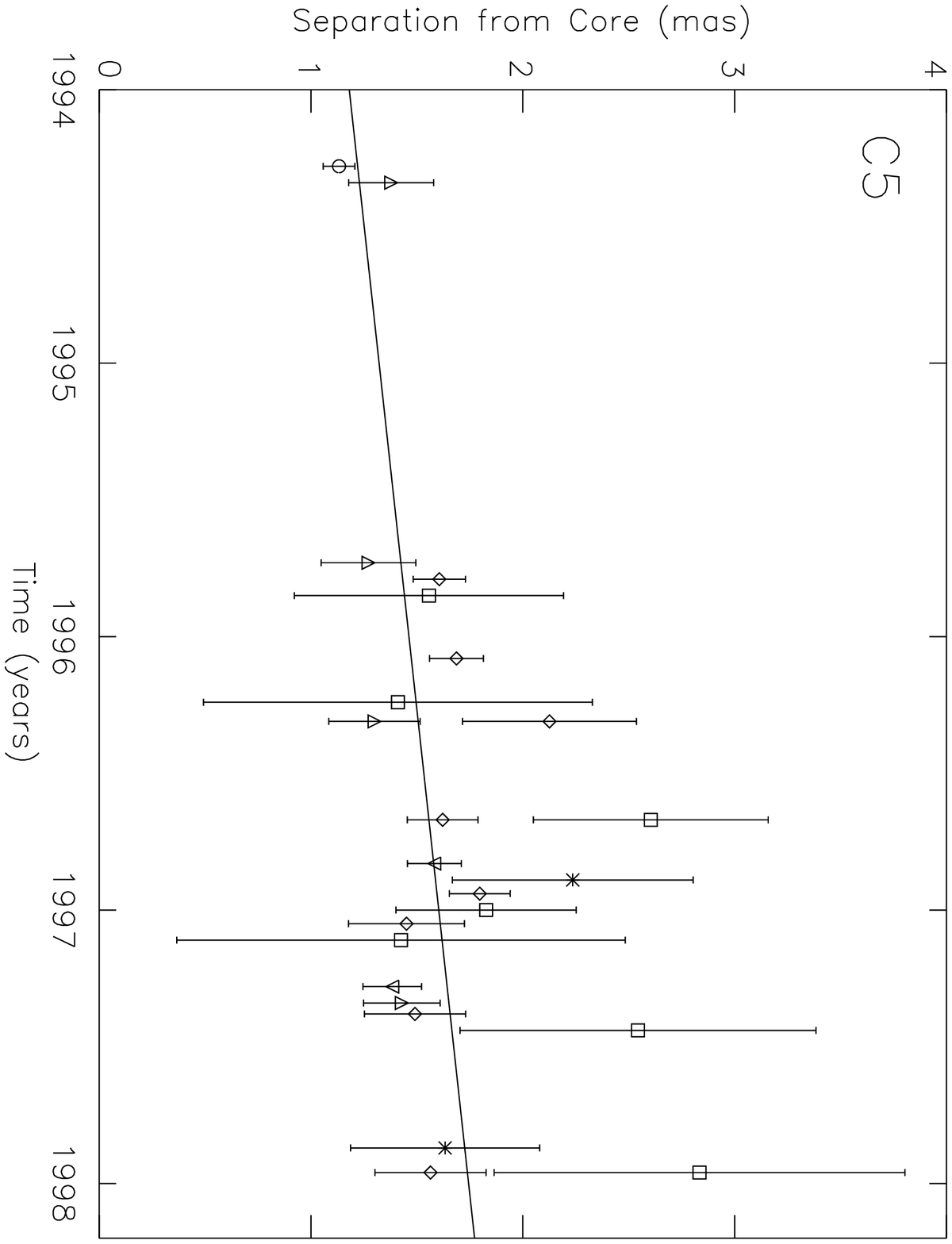}{0.0in}{90}{50}{50}{368}{214}
\plotfiddle{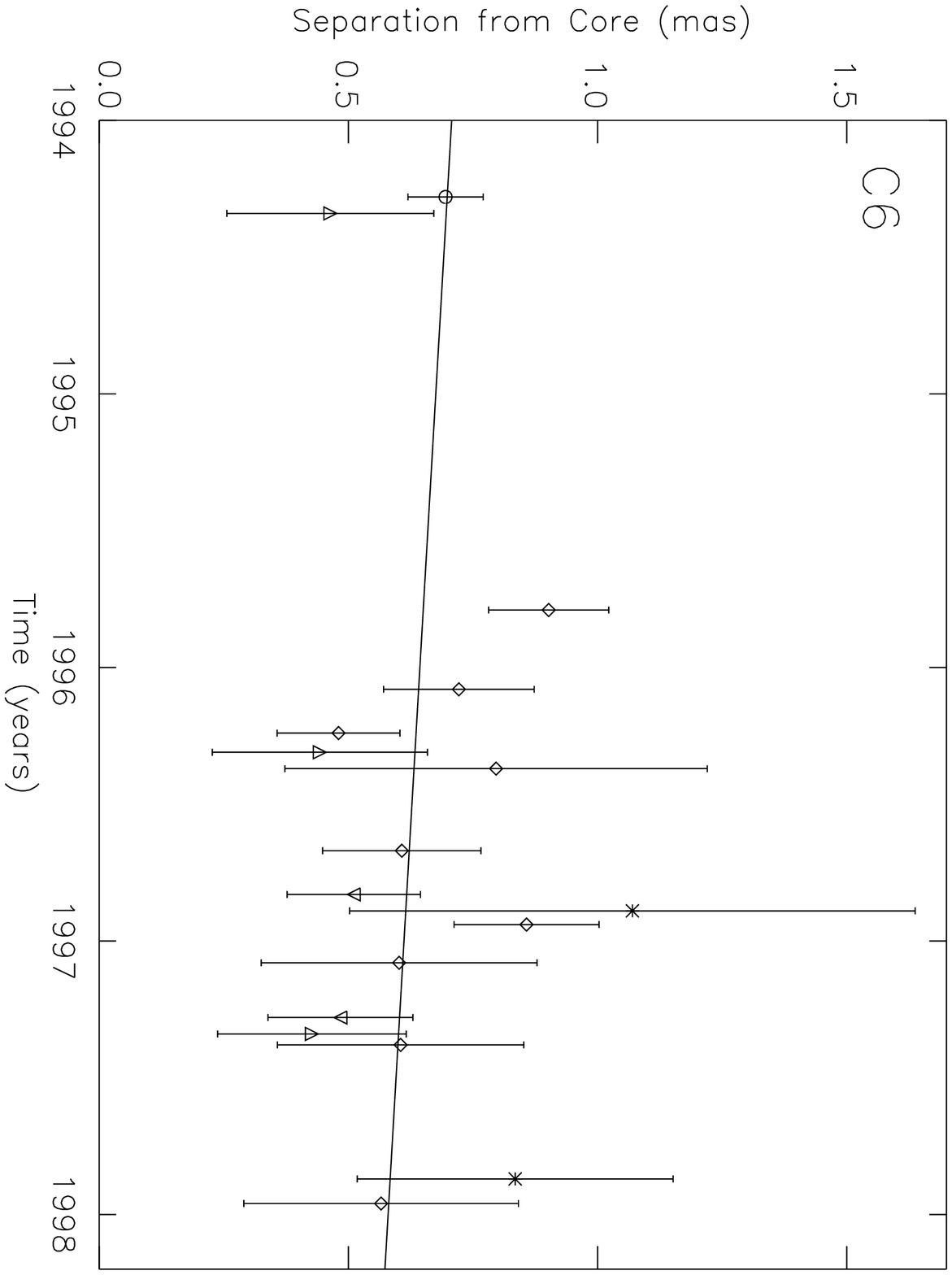}{0.0in}{90}{50}{50}{29}{-18}
\figcaption{($a$)-($c$): Distances from the core of components C4-C6 as
a function of time.  Squares represent the 2 GHz positions, asterisks 5 GHz,
diamonds 8 GHz, upward-pointing triangles 15 GHz,
downward-pointing triangles 22 GHz, and circles 43 GHz.
The lines shown are the best fits to motion with constant speed,
fitting the data from all frequencies together.
In cases where a component was observed at different frequencies at the same epoch,
a small shift was applied to the times of the plotted positions
(but not the fitted positions) at different frequencies to avoid
overlapping error bars.
($a$): C4  ($b$): C5  ($c$): C6.}
\end{sidewaysfigure*}

In order to determine the speeds of the components, we made linear least-squares fits to the separations of the
components from the core with time.  Initially, we fit the measurements made at each frequency
separately to avoid any effects from frequency-dependent separation ---
an effect which causes component positions measured at lower frequencies to appear closer to
the core (e.g. Biretta, Moore, \& Cohen 1986; Unwin et al. 1994; Piner \& Kingham 1997b; 
Lobanov 1998), and which may be due to gradients in magnetic field and electron density, such that
the $\tau =1$ surface moves progressively inward at higher frequencies.  All speed measurements in
this case were consistent (within 2$\sigma$) with no motion.  However, we observed that the average
separations were significantly different for the measurements at different frequencies for
C4 and C5, and in the opposite sense from the usually observed frequency-dependent separation
(i.e. the higher-frequency positions were on average closer to the core).  Since the higher-frequency
43 and 15 GHz observations occur earlier in the observed time range, we can interpret this as evidence
of motion and fit the data using all frequencies together.  These fits are shown as solid lines
on Figures 7$a$-$c$.  The speeds obtained from these fits are 0.19$\pm$0.27,
0.30$\pm$0.07, and $-$0.07$\pm$0.07~$c$ for C4 to C6 respectively.  If there is actual
frequency-dependent separation then these speeds are lower limits.  The possible magnitude
of this effect was investigated by assuming a power-law form for the frequency-dependent separation
(K\"{o}nigl 1981), and finding the maximum separation that still gave a
fit consistent with a linear fit at the 2$\sigma$ level.  The speed upper limits
obtained in this fashion were 0.20, 0.34, and $-$0.03~$c$ for C4 to C6 respectively.
Component C5 moves at subluminal speeds, although some care must be taken with this interpretation since
the measured speed is due solely to the earliest 43 GHz point; without this point the measured speed
is 0.05$\pm$0.12~$c$.  In fact, our data are consistent with no motion at all in any component,
over the time range from late 1995 to late 1997.

The position angles and their associated errors listed in Table~\ref{mfittab} do not show
any changes in the position angles of individual components, if we disregard
the position angle measurements at 43 GHz which are probably measured relative to a different point
since the core region is deconvolved into several components at this frequency.
Note that the position angle errors are relatively larger for the inner components, because
a given linear error translates into a larger angular error close to the core.
We do see a monotonic change in the position angle of the jet from component to component.  
Measured position angles moving outward along the jet are $-34\pm2\arcdeg$ for the average of C5 and C6,
$-38\pm1\arcdeg$ for the average of C3 and C4, $-46\pm2\arcdeg$ for C2, 
and $-64\pm3\arcdeg$ for C1.
Beyond component C1 the jet appears to bend back in the other direction; the diffuse emission seen
on two occasions to the northwest of C1 (Table~\ref{mfittab}) has a position angle of 
approximately $-55\arcdeg$.

Measurement of subluminal speeds, or no motion, instead of superluminal motion, 
in Mrk 421 is somewhat surprising given the high Doppler
factor required by the TeV variability.  In particular, care must be taken that
many rapidly moving components have not been misidentified as a few slower moving ones.  At the close distance
of Mrk 421 (z=0.03), an apparent speed of 10~$c$ would correspond to a proper motion of about 5 mas/yr.
We have tried to fit such rapidly moving components
to the data, but the dense time coverage provided by our observations and the
large number of measured positions preclude any such interpretation.
The only consistent interpretation including all model-fit component positions is in terms
of the subluminal components discussed above.

The component speeds measured in this paper differ from those measured by ZB90 in the early 1980's.
They measured  an apparent superluminal motion of two
jet components with a speed of about 2.9~$c$ (adjusted to our $H_{0}$) from 3 epochs of data at 5 GHz.
However, they were forced to reject the model fit to 
an earlier, fourth epoch of data in order to obtain a reasonable
fit with moving components, and some of the identified components (e.g. the last measurement
of their C1) were as faint as 1 mJy.  Let us instead posit that the
error bars on the ZB90
position measurements may be larger than the quoted values, which are 1\% of the 1 mas beam size for 
their component 2.  We believe this may be a reasonable assumption, especially given the
faintness of the fitted jet components.
Then the four entries for component 2 in their Table 1 may all be the same component.  The same
could hold for components 3 and 4 in their Table 1 (the non-detection of component 4 at their
second and third epochs could simply be due to the component's faintness), 
and this would remove the necessity for proceeding to
the evolutionary model in their Table 2.  This would also remove the necessity for rejecting the
first epoch model-fit.  The motion of these components would then be consistent with subluminal motion.

Accepting this interpretation of the ZB90 data, it is interesting to see if their
measured component positions correspond to an extrapolation of any of our measured component motions.
The ten-year gap between the ZB90 data and our data makes it impossible to prove any
such correlation, but it remains an interesting exercise.  The most likely pairing of components 
matches ZB90's Component 2 from their Table 1 with our C4, 
and ZB90's Component 3 from their Table 1 with our C3.
Assigning half-beam error bars to the ZB90 position measurements, we obtain fitted
speeds of 0.77$\pm$0.08~$c$ for C3 and 0.47$\pm$0.04~$c$ for C4, an increase of about 1$\sigma$
from the previously quoted speed of C4, assuming no deceleration over the intervening ten years.  
Note that an alternate interpretation that pairs
ZB90's Component 2 from their Table 1 with our C5, ZB90's Component 3 from their Table 1 with our C4, and
ZB90's Component 4 from their Table 1 with our C3 produces an unacceptable fit to the motion of C5.
ZB91 present two 22 GHz images of Mrk 421 from 1982 and 1984.  They detect no components
except for the core in their 1982 image, and detect a single component 0.88 mas from the core at
position angle $-72.5\arcdeg$ in their 1984 image.  This component is at a separation from the core
where it could conceivably match up with our C5 or our C6, but its position angle is approximately
40$\arcdeg$ from our measured position angle at this radius.  Even allowing half-beam error bars on
ZB91's measured angle ($\pm13.3\arcdeg$) the component has a position angle that differs by 3$\sigma$
from our measured position angles of C5 and C6.  If real, this component requires a sizable shift
in the position angle of the jet within 1 mas of the core in the ten years between the ZB91 image
and our images.

Marscher (1999) measures a speed of 2$c$ for Mrk 421 based on two 15 GHz images two weeks apart
from 1997 July 30 and 1997 August 15; however, he also mentions that an alternative interpretation
of his entire set of data spanning the three years from 1995 to 1998 is that the motion is slower, or about
0.3$c$.  This second interpretation yields a speed identical to that measured for C5 in this paper.

\section {Discussion}
\label{discussion}
\subsection{Subluminal Speeds}
\subsubsection{Small Angle to the Line-of-Sight}
Variability of the TeV emission from Mrk 421 on short timescales requires that the Doppler factor
in the TeV emitting region be high in order to avoid attenuation of the $\gamma$-rays by pair-production.
Gaidos et al. (1996) derive a Doppler factor $\delta\geq 9$ for a variability time of 30 minutes.
The lower limit to the Doppler factor derived from compactness arguments
is only weakly dependent on the variability time, $\delta\propto t^{-0.19}$ for Mrk 421 
(Gaidos et al. 1996).  Other authors have derived higher limits to the Doppler factor of Mrk 421 using the observed
TeV variability timescales and specific homogeneous synchrotron self-Compton (SSC) models.  
Mastichiadis \& Kirk (1997) derive $\delta=15(t/10^{5})^{-0.25}$, which gives $\delta\approx 40$
for the same variability time used by Gaidos et al. (1996).  Bednarek \& Protheroe (1997) derive an
allowed range of $\delta$ between 25 and 40 for a variability time of 15 minutes.  

If the apparent component speeds and the Doppler factor are both known, and the pattern speed seen
in VLBI images equals the bulk flow speed, then the angle of the jet to the
line-of-sight $\theta$ and the Lorentz factor $\Gamma$ can be calculated from 
\[\Gamma=\frac{\beta_{app}^{2}+\delta^{2}+1}{2\delta}\] and
\[\tan\theta=\frac{2\beta_{app}}{\beta_{app}^{2}+\delta^{2}-1},\] where $\beta_{app}$ is the 
apparent speed (Ghisellini et al. 1993).  
Because the applicability of the homogeneous SSC model to Mrk 421 is far from certain,
e.g. Bednarek \& Protheroe (1997), we take $\delta$ to be at the lower limit derived from
compactness arguments, $\delta=9$.  If we take the apparent speed to be the 
measured speed of C5, or 0.30~$c$
--- excluding C6 as a stationary component (such as those discussed by Vermeulen \& Cohen [1994]) ---
we obtain $\Gamma=4.6$ and $\theta=0.4\arcdeg$.  In this case, the subluminal motion arises because
the source is aligned almost exactly with the line-of-sight.
Simulations of statistical properties of gamma-ray loud blazars by Lister (1999a; 1999b)
indicate that such a small angle to the line-of-sight may not be that unlikely for a
gamma-ray loud blazar ---  approximately 15\% of his simulated sample lies within 0.5$\arcdeg$
of the line-of-sight, with the exact percentage depending on the specific $\gamma$-ray emission
model assumed.  However, these simulations also show that apparent subluminal motion is
considerably less likely, with at most 4\% of gamma-ray loud blazars being expected to
have observed speeds less than 1 $h^{-1}c$.
A small viewing angle also has implications for the source size and the jet opening angle.
Taking the projected angular size of Mrk 421 from the VLA image of Laurent-Muehleisen et al. (1993),
and assuming a viewing angle of 0.4$\arcdeg$, implies a rather large linear size of $\sim$1.5 Mpc,
although it is unlikely that the viewing angle would remain constant over these distances;
jet bending is evident in the VLBI images beyond about 10 mas from the core.  The opening angle
of the jet is also constrained to be less than the viewing angle, otherwise a unidirectional jet
would not be observed.  

\subsubsection{Changing Doppler Factor}
There are two other explanations for the observed subluminal speeds in Mrk 421 besides a very small viewing
angle.  The Doppler factor could change between the TeV emitting 
region at $\sim10^{-4}$ pc and the region where speeds are measured with VLBI  
at $\sim10$ pc (deprojected), or the speeds measured in VLBI observations may only reflect the motion of a 
pattern on an underlying flow of different speed.  We will discuss each of these possibilities in turn.
For the case of a changing Doppler factor, a change in viewing angle alone
is not sufficient to cause the slow observed speeds, 
even if the jet has its minimum Lorentz factor of $\Gamma=4.6$
(the minimum Lorentz factor for a jet with $\delta=9$).
A change in the Lorentz factor of the jet is also required.  If the jet maintains a constant angle 
to the line-of-sight of $5\arcdeg$ (a typical value for a jet with $\delta=9$), 
then a change in the bulk Lorentz factor from $\Gamma>5.6$
in the TeV producing region (where $\delta\geq9$) to $\Gamma=1.6$ in the VLBI jet 
(where $\beta_{app}=0.30$) is required.  Marscher (1999) offers an explanation for the deceleration
of jets in TeV blazars.  If the electron spectrum has a slope flatter than $-$2 --- as indicated by
the flat synchrotron spectra --- then most of the kinetic energy is in high energy electrons
that lose energy very efficiently to synchrotron radiation and inverse Compton scattering.
If the energy of the flow is put into particles via shocks, and the jet is composed mainly of
electrons and positrons, then most of the energy and momentum can be lost close to the 
base of the jet.

Such a change in the Doppler factor may be supported by the
absence of strong beaming indicators in wavebands other than TeV $\gamma$-rays.
As discussed in $\S$~\ref{intro}, the lack of variability in the EGRET light curve provides
no evidence for relativistic beaming of the GeV $\gamma$-rays.  
We have calculated the brightness temperature for each of the
non-delta function core component model fits in
Table~\ref{mfittab}, and we find that the measured brightness temperatures
also do not require Mrk 421 to be highly beamed.
We calculate the maximum brightness temperatures of the Gaussian core components using
\[T_{b}=1.22\times10^{12}\;\frac{S(1+z)}{ab\nu^{2}}\;\rm{K}\] (Tingay et al. 1998b),
where $S$ is the flux density of the component in Janskys, 
$a$ and $b$ are the FWHMs of the major and minor axes respectively in mas,
$\nu$ is the observation frequency in GHz, and $z$ is the redshift.  The median brightness temperature  
of our core model fits is 3$\times10^{11}$ K.  This value 
only requires a Doppler factor $\delta>3$ to reduce the intrinsic brightness temperature
$T_{b}/\delta$ below the brightness
temperature limit of $\sim10^{11}$ K derived by Readhead (1994) from equipartition arguments.
Brightness temperatures of 6$\times10^{10}$ and 8$\times10^{9}$ K have also been measured for Mrk 421 by
G\"{u}ijosa \& Daly (1996) and Kellermann et al. (1998) respectively.  These measurements are
below both the inverse Compton limit and the nominal equipartition brightness temperature limit.

Calculations in the literature of the equipartition and inverse Compton Doppler factors 
(from X-ray and VLBI data) of Mrk 421
(Ghisellini et al. 1993; G\"{u}ijosa \& Daly 1996; Guerra \& Daly 1997; Jiang, Cao, \& Hong 1998) 
have also all resulted in rather low estimates of $\delta$ --- all less than about 2.
However, there are certainly some problems with these estimates.  Some of these calculations
(Ghisellini et al. 1993; G\"{u}ijosa \& Daly 1996; Guerra \& Daly 1997)
may be inaccurate because they assume a homogeneous spherical source which is inconsistent with
the dependence of core size on observing frequency seen in Table~\ref{mfittab}, and because they assume
the synchrotron turnover frequency to be at the VLBI observing frequency when 
extrapolations of the radio and optical-IR spectra suggest it actually
occurs at around 10$^{12}$ Hz in this source (Makino et al. 1987).
The inverse Compton calculations --- including the inhomogeneous calculations by Jiang, Cao, \& Hong (1998)
--- are also inappropriate because they assume the X-ray emission from this source is inverse Compton,
when in fact it is synchrotron emission (Macomb et al. 1995, 1996; Takahashi et al. 1996).

The lack of strong beaming indicators in the radio and lower energy $\gamma$-rays is suggestive
of a jet where the lower frequency components of both the synchrotron and the inverse Compton
spectra (radio to optical/UV and Gev $\gamma$-rays respectively in the case of Mrk 421) 
are produced farther out in a region of lower Doppler factor than the higher frequency
components of these spectra (X-rays and TeV $\gamma$-rays in the case of Mrk 421).  
Maraschi, Ghisellini, \& Celotti (1992) have used a model where the high-frequency components
of the synchrotron and inverse Compton spectra are produced in a region closer to the core than the
low-frequency components to explain the GeV $\gamma$-ray emission of 3C279.

\subsubsection{Unequal Pattern and Bulk Flow Speeds}
The other possible explanation for the slow apparent speeds in Mrk 421 is that the pattern speed
seen in the VLBI images does not equal the bulk flow speed.  
If this is the case, then different Lorentz factors must be
used in the equations for $\beta_{app}$ and $\delta$, and they can no longer be solved simultaneously
for the angle to the line-of-sight.  The assumption of equal bulk and pattern speeds
is often made for the sake of simplicity, but has little to justify it.  In fact, Vermeulen \& Cohen (1994)
show that ratios of pattern to bulk Lorentz factors ranging from $\sim0.5$ to
$\sim5$ can fit the observed $\beta_{app}$ distribution of core-selected quasars.  
Tingay et al. (1998a) noted that the rapid internal
evolution of VLBI components in the jet of Centaurus A, when compared with the slow apparent speeds,
implies that the pattern speed is significantly less than the bulk flow speed in this source.
Hydrodynamical simulations by G\'{o}mez et al. (1997) give the result that the maximum apparent speed
corresponds to the bulk Lorentz factor, but that slower speeds can also be observed.
Stationary components have been observed to coexist with superluminal components in several sources
(e.g. 4C39.25 [Alberdi et al. 1993] and 1606+106 [Piner \& Kingham 1998]),
and Vermeulen \& Cohen (1994) remark that the large number of stationary components seen in
core-dominated quasars precludes them being part of a continuous distribution of Lorentz factors.
Mrk 421 is the first case known to us where all VLBI components present
in a series of maps at many epochs and frequencies have been observed to be
stationary or subluminal in a source known from other evidence to be strongly beamed.
If we assume a Doppler factor of $\delta=9$ and a viewing angle of $5\arcdeg$ 
for Mrk 421, then we obtain a bulk Lorentz factor of $\Gamma_{b}=5.6$ and
a pattern Lorentz factor of $\Gamma_{p}=1.6$, or a ratio of 
pattern to bulk Lorentz factor of $\sim0.3$.

\subsection{Zero-Separation Epochs}
Regardless of the nature of the VLBI components, their motion can be extrapolated back to yield an
estimated zero-separation epoch when the component separated from the VLBI core, with the caveats
that the proper motion may not have been constant over this time and that the VLBI ``core'' is simply
the point at which the jet becomes optically thick to radiation of that frequency.  
These epochs can then be compared with the total flux density light curves.
It is expected that the zero-separation epochs of VLBI components
should correlate with the beginnings of flares in the radio light curves (e.g. Mutel et al. 1990; 
Zensus et al. 1990; Valtaoja et al. 1999), 
since newly emerging VLBI components typically contribute significantly to the total flux density of the source. 
Extrapolating back the proper motion of C5 yields a zero-separation epoch of
1985.6.  Comparing this to the UMRAO light curves, we find that this corresponds with the beginning
of a radio flare that peaked in 1989.  Two flares subsequent to this are evident in the UMRAO data,
one beginning in 1991, the other in 1996 (The structure of this flare at higher radio frequencies  
is discussed by Tosti et al. [1998]).  If C6 had a proper motion similar to C5, it would be at
about the right distance to have been ejected in 1991; however, C6 presently appears to be stationary. 
The zero-separation epoch of C4 is well before the beginning of monitoring of Mrk 421.

Correlations can also be looked for between the emergence of VLBI components and flares in
other wavebands.  The largest TeV $\gamma$-ray flare yet observed from Mrk 421 occurred in 1996 May 
(Gaidos et al. 1996).  The highest resolution VLBI observations after this flare are the two
22 GHz VLBA observations in 1996 November and 1997 May (epochs 8 and 11).  These two observations
both show a component between C6 and the core, at a distance of about 0.15 mas.  However, since
no apparent motion is evident in this component between these two images, and since a component at a 
similar distance is seen in the 43 GHz image from 1994 April, no correlation can be claimed between
this component and the 1996 May TeV flare.

\section{Conclusions}

We have presented 30 VLBI images of the BL Lac object Markarian 421 which show the parsec-scale
structure of the radio jet on scales from 0.5 to approximately 30 mas from the core
(0.3 to 20 pc projected linear distance) over the years 1994 to 1997 at frequencies from
2 to 43 GHz.  We measured the apparent speeds of three inner jet components and found them
all to be subluminal (0.19$\pm$0.27,
0.30$\pm$0.07, and $-$0.07$\pm$0.07~$c$ for C4 to C6 respectively), and all to be consistent
with no motion if the 43 GHz image is excluded.
The dense time coverage of these VLBI images --- with spacings from one image to the next
as short as 16 days --- allowed us to unambiguously track these components and rule out any
possible strobing effect which might be caused by more numerous rapidly moving components.
Subluminal motion in the parsec-scale radio jet most likely implies 
that the jet slows down between the TeV production region and the region imaged by these
VLBI observations, or that the pattern speed is much less than the fluid speed in this BL Lac.

As of this writing, only one of the four other blazars detected so far at TeV energies (see $\S$~\ref{intro})
has had the apparent speed of its jet components measured, and the measurements differ from each other.
Gabuzda et al. (1994) measure speeds in the range 0.6-1.2~$c$ for Mrk 501
by combining their two epochs of data at 5 GHz with a third epoch from Pearson \& Readhead (1988).
Giovannini et al. (1998) measure a speed of 5.2~$c$ for the same source, based on
two epochs of VSOP observations at 1.6 GHz, while Marscher (1999) measures a speed of 2.5$\pm0.3c$
from three epochs of 22 GHz VLBA data.  It would be interesting to see how
proper motion measurements for the other sources, together with Doppler factors estimated from TeV variability,
constrain the physics of the TeV blazars.

\acknowledgments
Part of the work described in this paper has been carried out at the Jet
Propulsion Laboratory, California Institute of Technology, under
contract with the National Aeronautics and Space Administration.
We thank Alan Marscher for communication of data in advance of publication,
and for his helpful comments as referee.
P.G.E. thanks George Moellenbrock for assistance with data analysis.
A.E.W. acknowledges support from the NASA Long Term Space Astrophysics Program.
We gratefully acknowledge the VSOP Project, which is led by the Japanese Institute of Space and
Astronautical Science in cooperation with many organizations and radio telescopes around the world.
The National Radio Astronomy Observatory is a facility of the National Science Foundation operated
under cooperative agreement by Associated Universities, Inc.
This research has made use of the United States Naval Observatory (USNO) Radio Reference Frame Image Database (RRFID),
data from the University of Michigan Radio Astronomy Observatory which is supported by
the National Science Foundation and by funds from the University of Michigan,
and the NASA/IPAC extragalactic database (NED)
which is operated by the Jet Propulsion Laboratory, California Institute of Technology, under contract
with the National Aeronautics and Space Administration.

\end{document}